\documentclass[12pt]{article}
\pdfoutput=1

\usepackage{draft}
\usepackage{dsfont}
\usepackage{bbm}

\newcommand{\Hom}{\text{Hom}}

\newcommand{\bO}[1]{\cO_{#1}}

\newcommand{\dO}[2]{\cO_{#1}^{#2}}

\newcommand{\CM}{\cM}

\newcommand{\tN}{\widetilde \cN}
\newcommand{\tF}{\widetilde F}

\newcommand{\td}{\widetilde d}
\newcommand{\BCO}{\cB}
\newcommand{\tC}{{\widetilde{\cC}}}
\renewcommand{\vev}[1]{\langle{#1}\rangle}
\newcommand{\vvev}[1]{\langle\!\langle{#1}\rangle\!\rangle}
\renewcommand{\cI}{\mathbbm{1}}
\renewcommand{\cN}{N}
\renewcommand{\cW}{W}
\newcommand{\one}{1}

\begin{document}

\begin{titlepage}

\preprint{CALT-TH 2021-035 \\
YITP-SB-2021-19
}

\begin{center}

\hfill \\

\title{
Construction of two-dimensional topological field theories with non-invertible symmetries
}

\author{Tzu-Chen Huang,$^{a}$ Ying-Hsuan Lin,$^{b}$ Sahand Seifnashri$^{c,d}$}

\address{${}^a$Walter Burke Institute for Theoretical Physics,
\\
California Institute of Technology, Pasadena, CA 91125, USA}

\address{${}^b$Jefferson Physical Laboratory, Harvard University, Cambridge, MA 02138, USA}

\address{${}^c$Simons Center for Geometry and Physics, Stony Brook University,\\
Stony Brook, NY 11794-3636, USA}

\address{${}^d$C.\ N.\ Yang Institute for Theoretical Physics, Stony Brook University,\\ Stony Brook, NY 11794-3840, USA}

\email{jimmy@caltech.edu, yhlin@fas.harvard.edu, \\ sahand.seifnashri@stonybrook.edu}

\end{center}

\vfill

\begin{abstract}
We construct the defining data of two-dimensional topological field theories (TFTs) enriched by non-invertible symmetries/topological defect lines.
Simple formulae for the three-point functions and the lasso two-point functions are derived, and crossing symmetry is proven.
The key ingredients are open-to-closed maps and a boundary crossing relation, by which we show that a diagonal basis exists in the defect Hilbert spaces.
We then introduce regular TFTs, provide their explicit constructions for the Fibonacci, Ising and Haagerup $\mathcal{H}_3$ fusion categories, and match our formulae with previous bootstrap results.
We end by explaining how non-regular TFTs are obtained from regular TFTs via generalized gauging.
\end{abstract}

\vfill

\end{titlepage}

\tableofcontents

\section{Introduction and summary}

Topological defects generalize the notion of finite group global symmetries to higher symmetries~\cite{Kapustin:2013uxa,Gaiotto:2014kfa} and non-invertible symmetries.
In two dimensions, non-invertible topological defect lines (TDLs)~\cite{Petkova:2000ip,Bachas:2004sy,Frohlich:2006ch,Davydov:2010rm,Bhardwaj:2017xup,Chang:2018iay} are captured by fusion categories, so such generalized symmetries are sometimes called fusion category symmetries \cite{Bhardwaj:2017xup,Chang:2018iay,Thorngren:2019iar}.
We will denote a fusion category by $\cC$ and use abbreviations such as $\cC$-symmetry.

Lately, there has been a burgeoning body of work \cite{Chang:2018iay,Thorngren:2019iar,Komargodski:2020mxz,Thorngren:2021yso,Kikuchi:2021qxz} exploring the constraints on renormalization group flows by fusion category symmetries.
On the lattice, non-invertible symmetries can be directly built into the construction of anyon chains \cite{Feiguin:2006ydp} and related statistical models \cite{Aasen:2016dop,Aasen:2020jwb}; in continuum field theory, symmetry-preserving flows are triggered by relevant operators that are invariant under $\cC$, i.e.\ commute with the TDLs in $\cC$.
In either case, the deep infrared in a gapped phase must be described by a $\cC$-symmetric topological field theory \cite{Bhardwaj:2017xup,Chang:2018iay} that incorporates the structure of the fusion category symmetry.

There is a vast literature \cite{dijkgraaf1989geometrical,Fukuma:1993hy,Durhuus:1993cq,Sawin:1995rh,Abrams:1996ty,kock2004frobenius,Lazaroiu:2000rk,Alexeevski:2002rp,Lauda:2005wn,Lauda:2006mn,Moore:2006dw,Runkel:2008gr,Davydov:2011kb,Bhardwaj:2017xup,Chang:2018iay,Thorngren:2019iar,Komargodski:2020mxz,Inamura:2021wuo} on two-dimensional topological field theories (TFTs) with varying amounts of structure.
Like other types of TFTs, $\cC$-symmetric TFTs can be formulated axiomatically, either in the language of transition amplitudes associated to cobordisms \cite{Bhardwaj:2017xup}, or in terms of the correlators of point-like operators \cite{Chang:2018iay}.
In the latter formulation, the \emph{defining data} of a closed $\cC$-symmetric TFT consists of a fusion category $\cC$, the spectra of point-like operators, their three-point functions, and a set of generalized $\cC$-actions on point-like operators called {\it lassos}.
The most general correlator can be built from such {defining data}, and as long as the crossing symmetry of the four-point functions and the modular invariance of the torus one-point functions are satisfied, different ways of building the same correlator give rise to the same result \cite{Sonoda:1988mf,Sonoda:1988fq,Moore:1988qv,Moore:1989vd,bakalov2000lego,Chang:2018iay}.

In \cite{Thorngren:2019iar}, it was argued that $\cC$-symmetric TFTs are in one-to-one correspondence with module categories over $\cC$.
On the one hand, they examined the consistency of the $\cC$-symmetry in the presence of boundaries.
On the other hand, they realized two-dimensional TFTs as three-dimensional Turaev-Viro \cite{Turaev:1992hq} theories on a thin strip, and studied the consistency at the boundary of the strip.
In both cases, they found the consistency to be equivalent to the axioms of a module category.
However, exactly how a module category determines the above \emph{defining data} of a $\cC$-symmetric TFT was not expounded, except in the special case of fiber-functors---TFTs with one-dimensional bulk Hilbert spaces.
Later in~\cite{Komargodski:2020mxz}, the bulk Frobenius algebra of local operators was constructed for any given module category, but the correlators of point-like defect operators were not considered.
In earlier work, constructions of TFTs were given for general invertible symmetries \cite{Wang:2017loc,Tachikawa:2017gyf}, and for concrete examples of non-invertible symmetries by bootstrap \cite{Chang:2018iay,Huang:2021ytb}.

The purpose of this paper is to provide the full construction of the defining data of a unitary/reflection-positive (and more generally semisimple) $\cC$-symmetric TFT for any given fusion category $\cC$ and any given module category.
The key formulae are summarized in section~\ref{sec:summary}.
It is remarkable that while module categories naturally describe the admissible boundary conditions in the open sector of a $\cC$-symmetric TFT, it also completely determines the closed sector, echoing the sentiments of \cite{Moore:2006dw} in the context of open/closed TFT.

\subsection{Overview of the construction}

A summary of the heart of this paper, section~\ref{Sec:Axiomatic}, is now presented.
In section~\ref{Sec:Hilbert}, we explain how the open sector of a $\cC$-symmetric TFT is already captured by the structure of a module category over $\cC$, and how the closed sector can be constructed by open-to-closed maps.
In section~\ref{Sec:Open}, we show that the module trace allows the computation of arbitrary correlators of boundary-changing operators on the disc.
We then move on to considering correlators in the closed sector.
In an open/closed TFT without $\cC$-symmetry, it was shown in \cite{Moore:2006dw} that simple boundary conditions under the open-to-closed map land on the diagonal basis (also called the projector or idempotent basis) of the bulk Frobenius algebra.
In the present $\cC$-symmetric case, we see that a very similar structure arises.
This structure can be understood intuitively via a boundary crossing relation, which we formulate in section~\ref{Sec:BoundaryCrossing}, both from a physical cutting-and-sewing perspective, and mathematically from the axioms of a module category with a module trace.
Sections~\ref{Sec:CWD} and~\ref{Sec:Crossing} delineate how the boundary crossing relation allows the straightforward derivation of compact formulae expressing the three-point functions (including the bulk Frobenius algebra) and the lassos purely in terms of the module categorical data.
In section~\ref{Sec:Crossing}, we mathematically prove that our formula for the four-point function, which manifests crossing symmetry, has the correct cutting-and-sewing decomposition into three-point functions.
A summary of our formulae can be found in section~\ref{sec:summary}.

The other sections are organized as follows.
In section~\ref{Sec:Category}, we review the fusion category of topological defect lines, the module category of boundary conditions, along with the trace and adjoint structures.
In section~\ref{Sec:Examples}, we introduce the notion of a regular $\cC$-symmetric TFT corresponding to the regular module category, apply our formulae to construct the regular TFTs for three concrete fusion categories, Fibonacci, Ising, and Haagerup $\cH_3$ \cite{haagerup1994principal,asaeda1999exotic,Grossman_2012}, and match with previous bootstrap constructions \cite{Chang:2018iay,Huang:2021ytb}.
In section~\ref{Sec:gauging}, we review the notion of generalized gauging, and explain how all non-regular TFTs can be obtained from gauging regular TFTs.

\section{Categories of topological defect lines and boundary conditions}
\label{Sec:Category}

In quantum field theory (QFT), in addition to local operators, there also exist extended objects such as codimension-one walls.
An example of such a wall is an interface separating two QFTs.
When the theories on the two sides of the interface are identical, we get a self-interface which we call a \emph{defect}.
And when the theory on one side is trivial, we get a \emph{boundary condition}.

One can consider the fusion of two defects by bringing them parallel and close together.
This defines a fusion structure on defects, similar to the operator product expansion (OPE) of local operators.
Moreover, defects can form junctions, and extended junctions can form junctions-of-junctions, and so on.
Putting all these structures together, defects form a kind of \emph{category} \cite{Baez:1995xq,Kapustin:2010ta} as opposed to an algebra like the OPE of local operators.
A special class of defects are \emph{topological} defects, which encompass and generalize the notion of finite global symmetries in QFT \cite{Frohlich:2006ch,Davydov:2010rm,Kapustin:2013uxa,Kapustin:2014gua,Gaiotto:2014kfa,Bhardwaj:2017xup,Chang:2018iay,Benini:2018reh}.

In two dimensions, topological defect lines (TDLs) when finitely generated (and under an additional condition) are mathematically described by {\it fusion categories} \cite{Etingof:aa,etingof2016tensor}.
The subject of this paper is on two-dimensional topological field theories (TFTs), where not only the defect lines, but also the boundaries are topological.
The goal of this section is to describe the mathematical structure behind these objects; in particular, we will explain the meaning of a category and its underlying data.

We mainly focus on unitary/reflection-positive TFTs, but our results are generally valid for \emph{semisimple} ones.
Relatedly, our discussion is valid not only for unitary fusion categories but for general fusion categories.
The notion of semisimplicity will be introduced later.

\subsection{Category of topological defect lines}

We first review the fusion category $\cC$ describing a finitely generated set of TDLs.
Since $\cC$ generalizes the notion of global symmetries, it is also called a fusion category symmetry \cite{Bhardwaj:2017xup,Chang:2018iay,Thorngren:2019iar}.

In a TFT, $\cC$ is different from the category of \emph{all} TDLs.
In the situation where the TFT is the endpoint of a symmetry-preserving RG flow, $\cC$ describes the symmetries not just at the endpoint, but along the RG trajectory.
On the one hand, the category of all TDLs in the TFT can be larger than $\cC$ since accidental symmetries can emerge.
On the other hand, some TDLs in $\cC$ can become identical at the endpoint, in which case $\cC$ does not act faithfully on the TFT.\footnote{The symmetry $\cC$ is sometimes called the extrinsic symmetry in the literature~\cite{Benini:2018reh}, while the category of all TDLs is called the intrinsic symmetry.}

Below we review the various structures of $\cC$.

\subsubsection*{Direct sum and simple objects}

Given any two TDLs, we can construct another TDL by taking the direct sum, an operation denoted by $\oplus$.
Any TDL that cannot be decomposed into a direct sum of more than one TDL is called a \emph{simple} (or indecomposable) TDL.
We use $a, b, c, \dotsc$ to label simple TDLs
\ie
\lu{$a$} \, ,
\fe
and denote the unit by $\cI$.
In the {categorical language}, they are the \emph{simple objects} of $\cC$.

Let us explain the physical meaning of the direct sum operation.
When two TDLs are supported on the same space-like curve $\gamma$, they become line \emph{operators} $U^a(\gamma)$ and $U^b(\gamma)$ acting on the Hilbert space $\mathcal{H}(\gamma)$.
In this case, $U^{a \oplus b}(\gamma) = U^a(\gamma) + U^b(\gamma)$.
But when the TDLs are time-like, they impose twisted boundary conditions and define defect(-twisted) Hilbert spaces $\mathcal{H}^a$ and $\mathcal{H}^b$, and we get $\mathcal{H}^{a \oplus b} = \mathcal{H}^a \oplus \mathcal{H}^b$.
More generally, a correlator involving $a \oplus b$ is the sum of the same correlator involving only $a$ and that involving only $b$.

Note that there is no notion of taking non-integral linear combinations of TDLs.
Such linear combinations lead to ill-defined defect Hilbert spaces with non-integral dimensions.
Such is the reason that the space of TDLs is a category, and not a vector space.
The same remark applies to the boundary conditions introduced later.

\subsubsection*{Junctions and morphisms}

We use $x, y, z, \dotsc$ to denote the junction vectors of TDLs, which in the {categorical language} correspond to the morphisms in $\cC$.
For instance, a junction vector between TDLs $a$ and $b$ corresponds to a morphism $x \in \mathrm{Hom}_\cC(a,b)$
\ie
\begin{gathered}
\begin{tikzpicture}[scale=1]
\draw [line,->-=.55] (0,-1) -- (0,0);
\draw (0,-0.6) node [right] {$a$};
\draw [line,->-=.55] (0,0) -- (0,1);
\draw (0,0.6) node [right] {$b$};
\filldraw[black] (0,0) circle (1pt) node[right] {\scriptsize $x$};
\end{tikzpicture}
\end{gathered} \, .
\fe
Given any two junction vectors $x \in \mathrm{Hom}_\cC(a,b)$ and $y \in \mathrm{Hom}_\cC(b,c)$, we can fuse them to get a new junction vector between $a$ and $c$.

In the {categorical language}, this corresponds to composing morphisms, and the new junction vector is denoted by $y \circ x \in \mathrm{Hom}_\cC(a,c)$.
Under direct sum, $\mathrm{Hom}(a \oplus b,c) = \mathrm{Hom}(a,c) \oplus \mathrm{Hom}(b,c)$ and $\mathrm{Hom}(a,b \oplus c) = \mathrm{Hom}(a,b) \oplus \mathrm{Hom}(a,c)$.

Note the distinction between $\mathrm{Hom}_\cC(a,b)$ and the space of \emph{all} $a \to b$ line-changing operators in a TFT.
Whereas the former is a property of the symmetry alone, the latter is theory-dependent.
For instance, $\mathrm{Hom}_\cC(\cI,\cI)$ is always one-dimensional, but the space of all $\cI \to \cI$ line-changing operators certainly does not have to be one-dimensional.
This is related to the aforementioned fact that physically, the symmetry $\cC$ is not the category of all TDLs in the TFT, but rather the category of TDLs that are preserved on a nearby RG trajectory.
Mathematically speaking, there is a tensor functor from $\cC$ to all the TDLs of the theory, which means that there is a map between the objects and morphisms of $\cC$ into the full category of TDLs and line-changing operators of the theory, while preserving the fusion categorical structure of $\cC$.
Moreover, this map is not required to be injective or surjective.\footnote{If this functor is not injective, then the symmetry $\cC$ is not faithful.
Although a non-faithful symmetry is sometimes regarded as an unauthentic symmetry of a QFT, in the TFT context it is useful to consider non-faithful symmetries.
For instance, for any non-anomalous finite group $G$, one can endow the trivial TFT with a non-faithful symmetry $G$, and gauge $G$ to produce a nontrivial TFT with $\text{Rep}(G)$ symmetry \cite{Bhardwaj:2017xup}.
}

We have now introduced enough categorical structure to explain the meaning of semisimplicity.
A semisimple category is one in which the notions of simplicity---an object $a$ satisfies $\Hom(a,a) \cong \bC$---and indecomposability---$a$ cannot be written as the direct sum of two objects---are equivalent.
In Appendix \ref{app.semisimple} we show that in a unitary/reflection-positive TFT, both the category of TDLs and the category of boundary conditions are semisimple.\footnote{Note that a semisimple tensor category with finitely many simple objects is by definition a multifusion category.
If the unit is also simple, then it is by definition a fusion category.
}
Throughout this paper we conflate simplicity and indecomposability.

\subsubsection*{Fusion rules, junctions and F-moves}

Fusion rules describe the fusion of parallel loops of TDLs on a flat cylinder
\ie
\begin{gathered}
\begin{tikzpicture}[scale=.5]
\draw [bg] (0,4) ++ (0:2 and 1) arc (0:180:2 and 1);
\draw [bg] (0,4) ++ (180:2 and 1) arc (180:360:2 and 1);
\draw [bg] (-2,0) -- (-2,4);
\draw [bg] (2,0) -- (2,4);
\draw [bg] (0,0) ++ (180:2 and 1) arc (180:360:2 and 1);
\draw [line,->-=.3] (0,1) ++ (180:2 and 1) arc (180:270:2 and 1) node[below] {$b$} arc (270:360:2 and 1);
\draw [line,dashed] (0,1) ++ (0:2 and 1) arc (0:180:2 and 1);
\draw [line,->-=.3] (0,2) ++ (180:2 and 1) arc (180:270:2 and 1) node[above] {$a$} arc (270:360:2 and 1);
\draw [line,dashed] (0,2) ++ (0:2 and 1) arc (0:180:2 and 1);
\end{tikzpicture}
\end{gathered}\quad \to \quad
\begin{gathered}
\begin{tikzpicture}[scale=.5]
\draw [bg] (0,4) ++ (0:2 and 1) arc (0:180:2 and 1);
\draw [bg] (0,4) ++ (180:2 and 1) arc (180:360:2 and 1);
\draw [bg] (-2,0) -- (-2,4);
\draw [bg] (2,0) -- (2,4);
\draw [bg] (0,0) ++ (180:2 and 1) arc (180:360:2 and 1);
\draw [line,->-=.3] (0,1.5) ++ (180:2 and 1) arc (180:270:2 and 1) node[below] {$a \otimes b$} arc (270:360:2 and 1);
\draw [line,dashed] (0,1.5) ++ (0:2 and 1) arc (0:180:2 and 1);
\end{tikzpicture}
\end{gathered}~.
\fe
We denote the fusion rule of the TDLs in $\cC$ by
\ie
a \otimes b = \bigoplus_c \cN_{ab}^c \, c \, , 
\label{fusion.rules}
\fe
where the non-negative integers $N_{ab}^c$ are the fusion coefficients.

When TDLs form junctions among themselves, each junction is endowed with a junction vector space.
The fusion coefficient $N_{ab}^c$ counts the dimensionality of the vector space $V_{ab}^c = \mathrm{Hom}_\cC(a \otimes b,c)$ at the three-way junction of $a$, $b$ and $c$.
We emphasize again that the junction vector space $V_{ab}^c$ can be smaller than the full defect Hilbert space $\mathcal{H}^{a\otimes b \otimes \bar c}$.
A gauge of $\cC$ is a choice of bases for all junction vector spaces.

When multiple junctions are joined together, the splitting-and-joining isomorphisms are captured by the F-moves\footnote{We assume that the cyclic-permutation map captured by $F^{a, b, c}_\cI$ is trivial, and ignore the specification of cyclic ordering at each trivalent junction.
The generalization to nontrivial cyclic-permutation is straightforward.
}
\ie
\label{F}
\begin{gathered}
\begin{tikzpicture}[scale=1]
\draw [line,->-=.56] (-1,0) node [left] {\scriptsize $x$}
-- (-.5,0) node [above] {$e$} -- (0,0);
\draw [line,-<-=.56] (-1,0) -- (-1.5,.87) node [above left=-3pt] {$a$};
\draw [line,-<-=.56] (-1,0) -- (-1.5,-.87) node [below left=-3pt] {$b$};
\draw [line,-<-=.56] (0,0) -- (.5,-.87) node [below right=-3pt] {$c$};
\draw [line,->-=.56] (0,0) node [right] {\scriptsize $y$}
-- (.5,.87) node [above right=-3pt] {$d$};
\end{tikzpicture}
\end{gathered}
~~=~~
\sum_{f, z, w}
(F^{a, b, c}_{d; e, f})_{x, y}^{z, w}
\begin{gathered}
\begin{tikzpicture}[scale=1]
\draw [line,->-=.6] (0,-1) node [below] {\scriptsize $z$}
-- (0,-.5) node [right] {$f$} -- (0,0);
\draw [line,-<-=.56] (0,-1) -- (.87,-1.5) node [below right=-3pt] {$c$};
\draw [line,-<-=.56] (0,-1) -- (-.87,-1.5) node [below left=-3pt] {$b$};
\draw [line,-<-=.56] (0,0) -- (-.87,.5) node [above left=-3pt] {$a$};
\draw [line,->-=.56] (0,0) node [above] {\scriptsize $w$}
-- (.87,.5) node [above right=-3pt] {$d$};
\end{tikzpicture}
\end{gathered}
\, .
\fe
In other words, a four-way junction of TDLs can be built out of two three-way junctions in two different ways, and the F-move characterizes the relation between them.
The consistency of five-way junctions then imposes a constraint on the F-moves called the Pentagon identity, which when satisfied, the consistency of arbitrary-way junctions follows\cite{MacLane:1963unh}.

The fusion coefficients $N$ and the F-symbols $F$ capture the full data of the fusion category $\cC$.
When the TDLs in $\cC$ are invertible, $\cC$ describes an ordinary finite group symmetry, and the fusion rule and the F-moves reduce to the group law and the 't Hooft anomaly, respectively.

\subsubsection*{Dual structure and folding}

The dual, or orientation-reversal,\footnote{For junction vectors, the dual and the orientation-reversal refer to different maps, but for TDLs, the two notions are interchangeable.} of a simple TDL $a$ is denoted by $\bar a$, such that
\ie
\lu{$\bar a$} ~ := \quad \ld{$a$} \, .
\fe
Note that $\bar{\bar{a}} = a$.\footnote{In the categorical language, the identification between $a$ and $\bar{\bar{a}}$ is given by a pivotal structure which we assume exists.}  An algebraic characterization of $\bar a$ will also be given momentarily.
A dual structure can also be defined on junctions: given a junction vector $x\in \mathrm{Hom}(a,b)$, we can rotate the junction by 180 degrees to get the dual junction vector $x^\vee \in \mathrm{Hom}(\bar{b},\bar{a})$.\footnote{Note that in most of the main text, we denote $x^\vee$ simply as $x$.
When $\mathrm{Hom}(a,b) \neq \mathrm{Hom}(\bar{b},\bar{a})$, their distinction is clear from the context.
However, in the case of $b = \bar{a} \simeq a$, we have $\mathrm{Hom}(a,b)=\mathrm{Hom}(\bar{b},\bar{a}) \neq 0$, and we should be careful with differentiating $x$ and $x^\vee$. This subtlety is addressed in appendix~\ref{App:FS}.
}
More precisely, there exists a linear map
\begin{equation}
    \vee: \quad \mathrm{Hom}_\cC(a, b) \to \mathrm{Hom}_\cC(\bar b , \bar a) \, .
\end{equation}

Given a junction $x \in \mathrm{Hom}(a,b)$, we can fold it in four different ways to get junction vectors in $\mathrm{Hom}(a \otimes \bar{b},\cI)$, $\mathrm{Hom}(\bar{b} \otimes a,\cI)$, $\mathrm{Hom}(\cI,\bar{a} \otimes b)$, and $\mathrm{Hom}(\cI, b \otimes \bar{a})$.
In particular, starting from the identity junction $1_a \in \mathrm{Hom}(a,a)$, we can obtain a junction between $a\otimes \bar{a}$ and the unit $\cI$.
Conversely, we can unfold a junction $a\otimes \bar{a} \to \cI$ into a junction between $a$ and itself.
Hence we conclude that $N_{a \bar{a}}^\cI = N_{a \cI}^a = N_{\cI a}^a = 1$, where we used $a = a \otimes \cI = \cI \otimes a$.
We can thus characterize $\bar a$ algebraically as the TDL whose fusion with $a$ contains $\cI$.

A TDL $a$ is called {\it self-dual} if $a \simeq \bar a$, meaning that there exists a topological junction connecting $a$ and $\bar a$.
This junction cannot be trivialized if the so-called Frobenius-Schur indicator~\cite{fredenhagen1992superselection,ng2007higher} is nontrivial.
We assume trivial Frobenius-Schur indicator throughout the main text, and remark on the nontrivial case in appendix~\ref{App:FS}.

\subsubsection*{Quantum dimensions}

The quantum dimension of a TDL $a$ is defined as the expectation value of an empty loop,
\ie
d_a \equiv ~
\begin{gathered}
\begin{tikzpicture}[baseline=-2,scale=0.5]
\draw [line,->-=.025,->-=.525] (0,0) circle (1);
\draw (1,0) node [right] {$a$};
\end{tikzpicture}
\end{gathered} \, .
\fe
In more precise terms, shrinking the loop to a point produces a topological local operator proportional to the identity, and the proportionality constant is the quantum dimension $d_a$.\footnote{The reader may wonder why the topological local operator must be proportional to the identity and cannot be some other topological local operator when the TFT has degenerate vacua.
Recall that when we regard the TFT as the infrared limit of a renormalization group (RG) flow, $\cC$ does not describe the full set of TDLs in the TFT, but only the subset that is preserved on the RG trajectory.
If we assume that the QFT away from the TFT point has the identity as its unique topological local operator, then a loop of TDL in $\cC$ must shrink to the identity.
By continuity, this is also true in the TFT.
If one prefers not to consider the TFT in the context of RG flows, then the above property follows from the fact that $\cC$ is defined to be a fusion category.
Mathematically, the full structure of a TFT including all the TDLs and the extra topological local operators is described by a multifusion category, where a loop does not need to shrink to the identity.
The statement that all loops of TDLs shrink to the one-dimensional $\cI \to \cI$ morphism follows from the fact that we are considering a fusion category, and not a multifusion category.
}

The quantum dimensions must cohere with fusion.
Given two parallel loops of TDLs, taking their fusion first and then shrinking them must give the same answer as shrinking them successively,
\ie
\begin{gathered}
\begin{tikzpicture}[scale=0.5]
\draw [line,->-=.025,->-=.525] (0,-2) circle (1);
\draw (1,-2) node [left] {$a$};
\draw [line,->-=.02,->-=.52] (0,-2) circle (1.5);
\draw (1.5,-2) node [right] {$b$};
\draw [line,->] (3,-1) -- (5,0);
\draw [line,->-=.025,->-=.525] (7,0) circle (1);
\draw (8,0) node [right] {$a \otimes b$};
\draw [line,->] (11,0) -- (13,0);
\draw [line] (16,0) node {$\sum_c \cN_{ab}^c \, d_c$};
\draw [line,->] (3,-3) -- (5,-4);
\draw [line] (7,-4) node {$d_a$};
\draw [line,->-=.02,->-=.52] (7,-4) circle (1);
\draw (8,-4) node [right] {$b$};
\draw [line,->] (11,-4) -- (13,-4);
\draw [line] (15,-4) node {$d_a d_b$};
\end{tikzpicture}
\end{gathered} \, ,
\fe
giving rise to the constraint
\ie
\label{Eigen}
\sum_c \cN_{ab}^c \, d_c = d_a d_b
\fe
saying that the quantum dimensions solve the abelianzied fusion rules.
Alternatively, we say that the vector $d$ is a simultaneous eigenvector, and at the same time the eigenvalues of the fusion coefficient matrices $N_a$.
This constraint generally has multiple solutions, among which the choice of $d$ is part of the fusion categorical data.
However, if $\cC$ is a unitary fusion category (see below), which must be the case in a unitary/reflection-positive QFT, then the choice of $d$ is fixed by the fusion coefficients: it is given by the unique Frobenius-Perron eigenvector of the fusion coefficient matrices.

Finally, we define the (global) dimension of $\cC$ to be\footnote{In condensed matter literature, $\mathcal{D}=\sqrt{\mathrm{dim}(\cC)}$ is called the total quantum dimension of $\cC$, which is the square root of the global dimension.}
\ie
\mathrm{dim}(\cC) \equiv \sum_a d_a^2 \, ,
\fe

\subsubsection*{Adjoint structure and orientation-reversal}

For a given junction vector $x$, there exists an adjoint junction $x^\dagger$ given by the orientation-reversal of the junction (orientation-reversal also changes every TDL label $a$ to $\bar a$)
\ie
    \begin{gathered}
    \begin{tikzpicture}[scale=1]
    \draw [line,->-=.77,->-=.27] (0,-1) -- (0,-.5) node [right] {$a$} -- (0,0) \dtb{right}{0}{\scriptsize $x$} -- (0,.5) node [right] {$b$} -- (0,1);
    \end{tikzpicture}
    \end{gathered}
\quad~\mapsto\qquad
    \begin{gathered}
    \begin{tikzpicture}[scale=1]
    \draw [line,->-=.77,->-=.27] (0,-1) -- (0,-.5) node [right] {$b$} -- (0,0) \dtb{right}{0}{\scriptsize $x^\dag$} -- (0,.5) node [right] {$a$} -- (0,1);
    \end{tikzpicture}
    \end{gathered} \, .
\fe
More precisely, there exists an antilinear map\footnote{Changing $b$ to $b \otimes c$ gives $\dagger: ~ \mathrm{Hom}_\cC(a, b \otimes c) \to \mathrm{Hom}_\cC(b \otimes c, a)$.}
\begin{equation}
    \dagger: \quad \mathrm{Hom}_\cC(a, b) \to \mathrm{Hom}_\cC(b, a) \, ,
\end{equation}
which defines a bilinear form on the junction vector spaces,
\ie
\label{Theta}
\vev{y^\dagger x} \quad \equiv \quad \Thj{c}{b}{a}{y^\dagger}{x}~, \qquad u,v \in \mathrm{Hom}_\cC(a, b \otimes c)~.
\fe
If this bilinear form is positive-definite, then the underlying fusion category is a unitary fusion category.\footnote{More precisely, this positive-definiteness defines a \emph{pseudo-}unitary fusion category; a unitary fusion category further requires that the F-symbols are unitary.
There is no proof that every pseudo-unitary fusion category is gauge-equivalent to a unitary one, but there is also no known counter-example \cite{etingof2016tensor}.
Therefore, we do not differentiate  between the two notions in this paper.
}
The TDLs in a unitary/reflection-positive QFT must be described by a unitary fusion category.

We are now ready to discuss boundary conditions.

\subsection{Category of boundary conditions}

Let us consider the category formed by \emph{all} boundary conditions (BCs) in a unitary/reflection-positive TFT with fusion category symmetry $\cC$, and denote it by $\CM$.
The fusion of TDLs with BCs defines an action of $\cC$ on $\cM$, thereby making $\cM$ a module category over $\cC$ \cite{Bhardwaj:2017xup}.

Below we review the various structures of $\cM$.

\subsubsection*{Simple objects and direct sum}
We use $m, n, k, \dotsc$ to label simple BCs
\ie
\begin{gathered}
\begin{tikzpicture}[scale=1]
\fill [fill=gray!20] (0,-1) rectangle (1,1);
\draw [line,->-=.55] (0,-1) -- (0,0) node [left] {$m$} -- (0,1);
\end{tikzpicture}
\end{gathered} \, , \label{boundaryline}
\fe
which correspond to the simple objects in $\cM$.
Our convention is that the empty theory, indicated by the gray filling, is always to the right of a boundary marked with an upwards arrow, and to the left of one marked with a downwards arrow.

Just like TDLs, we can construct non-simple BCs by taking direct sums of simple BCs.
When a BC $m$ is on a closed \emph{space-like} boundary, it corresponds to a boundary state $\ket{m}$
\ie
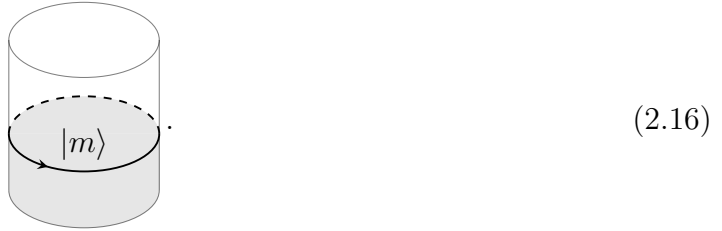
\begin{gathered}
\begin{tikzpicture}[scale=.5]
\fill [fill=gray!20] (0,1.5) ++ (0:2 and 1) arc (0:180:2 and 1);
\fill [fill=gray!20] (-2,0) rectangle (2,1.5);
\fill [fill=gray!20] (0,0) ++ (180:2 and 1) arc (180:360:2 and 1);
\draw [bg] (0,4) ++ (0:2 and 1) arc (0:180:2 and 1);
\draw [bg] (0,4) ++ (180:2 and 1) arc (180:360:2 and 1);
\draw [bg] (-2,0) -- (-2,4);
\draw [bg] (2,0) -- (2,4);
\draw [bg] (0,0) ++ (180:2 and 1) arc (180:360:2 and 1);
\draw [line,->-=.3] (0,1.5) ++ (180:2 and 1) arc (180:270:2 and 1) node[above] {$\ket{m}$} arc (270:360:2 and 1);
\draw [line,dashed] (0,1.5) ++ (0:2 and 1) arc (0:180:2 and 1);
\end{tikzpicture}
\end{gathered} \, .
\fe
Under direct sum, $\ket{m\oplus n} = \ket{m} + \ket{n}$.\footnote{Boundary states do not reside in a projective vector space like usual states in quantum mechanics do, so one cannot arbitrarily rescale their normalization.
}
When two parallel boundaries are \emph{time-like}, they define an open Hilbert space on the interval, denoted by $\mathcal{H}_{m,n}$
\ie
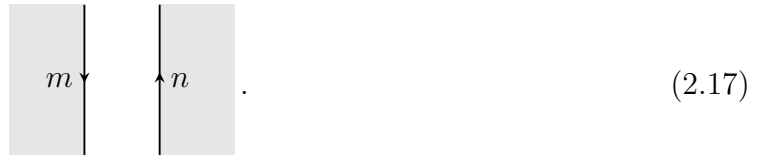
\begin{gathered}
\begin{tikzpicture}[scale=1]
\fill [fill=gray!20] (0,-1) rectangle (1,1);
\fill [fill=gray!20] (-2,-1) rectangle (-1,1);
\draw [line,-<-=.55] (-1,-1) -- (-1,0) node [left] {$m$} -- (-1,1);
\draw [line,->-=.55] (0,-1) -- (0,0) node [right] {$n$} -- (0,1);
\end{tikzpicture}
\end{gathered} \, .
\label{boundaryline}
\fe
Under direct sum, $\mathcal{H}_{m \oplus n,k} = \mathcal{H}_{m,k} \oplus \mathcal{H}_{n,k}$.

\subsubsection*{Boundary changing operators and morphisms}

A junction between two BCs $m$ and $n$ corresponds to a morphism $x \in \mathrm{Hom}_\cM(m, n)$ 
\ie
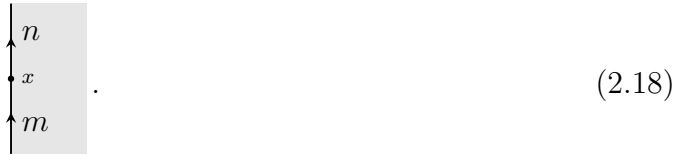
\begin{gathered}
\begin{tikzpicture}[scale=1]
\fill [fill=gray!20] (0,-1) rectangle (1,1);
\draw [line,->-=.55] (0,-1) -- (0,0);
\draw (0,-0.6) node [right] {$m$};
\draw [line,->-=.55] (0,0) -- (0,1);
\draw (0,0.6) node [right] {$n$};
\filldraw[black] (0,0) circle (1pt) node[right] {\scriptsize $x$};
\end{tikzpicture}
\end{gathered} \, .
\fe
Here, since $\cM$ is the category of \emph{all} boundary conditions, the above junction vector space is the same as the Hilbert space $\cH_{m,n}$ of all boundary-changing operators between $m$ and $n$,\footnote{By contrast, the junction vector spaces for TDLs in $\cC$ can be proper subspaces of defect Hilbert spaces.
}
\begin{equation}
    \mathrm{Hom}_\cM(m, n) = \mathcal{H}_{m,n} \, .
\end{equation}

The meaning of semisimplicity for $\cM$ is completely parallel to that for TDLs discussed earlier.
By assuming semisimplicity, any BC in $\cM$ is equivalent to a direct sum of simple ones $m$ with $\mathrm{Hom}_\cM(m, m) \cong \bC$.

\subsubsection*{Fusion rules, junctions and F-moves}

Similar to the fusion of TDLs among themselves, we denote the fusion of TDLs with BCs by
\ie
\label{BoundaryFusion}
a \otimes m = \bigoplus_{n} \tN_{am}^n \, n \, , \quad \tN_{am}^n \in \bZ_{\ge0} \, .
\fe
Each non-negative integer $\tN_{am}^n$ counts the dimension of the vector space $\mathrm{Hom}_\cM(m, a \otimes n)$ of all boundary-changing defect operators at the junction of $a$ with $m$ and $n$,
\ie
\begin{gathered}
\begin{tikzpicture}[scale=1]
\fill [fill=gray!20] (0,-1) rectangle (1,1);
\draw [line,->-=.55] (0,-1) -- (0,0);
\draw (0,-0.6) node [right] {$m$};
\draw [line,->-=.55] (0,0) -- (0,1);
\draw (0,0.6) node [right] {$n$};
\draw [line,->-=.55] (0,0) -- (-1,0);
\draw (-0.6,0) node [below] {$a$};
\filldraw[black] (0,0) circle (1pt) node[right] {\scriptsize $x$};
\end{tikzpicture}
\end{gathered} \,, \qquad x \in \mathrm{Hom}_\cM(m, a \otimes n) \, .
\fe
Under the operator-state map, this vector space is the same as the defect Hilbert space on an interval with BCs $m$ and $n$, and twisted by the TDL $a$ in between.
We denote this Hilbert space by $\cH_{m,n}^a$, and therefore
\begin{equation}
    \tN_{am}^n = \mathrm{dim} \, \mathcal{H}^a_{m,n}~.
\end{equation}

Since the action of TDLs on BCs is associative, the matrices $\widetilde{N}_a$ with $(\widetilde{N}_a)_{m,n} = \widetilde{N}_{am}^n$ furnish a \emph{nonnegative-integer-valued matrix representation (NIM-rep)} of the fusion algebra \eqref{fusion.rules},\footnote{In the context of rational conformal field theory, NIM-reps of the Verlinde fusion algebra~\cite{Verlinde:1988sn} have been studied in \cite{Cardy:1989ir,Behrend:1999bn,Gannon:2001ki,Gaberdiel:2002qa}.
}
\begin{equation}
    \sum_c N_{ab}^c \tN_{c} = \tN_{b} \tN_{a}~.
\end{equation} 
We fix a gauge for $\cM$ by choosing a basis for each Hilbert space $\mathcal{H}^a_{m,n}$.
The associator of the action of TDLs on BCs provides the F-moves for the action of $\cC$ on $\cM$
\ie
\label{BoundaryF}
\begin{gathered}
\begin{tikzpicture}[scale=1]
\fill[fill=gray!20] (0,0)--(.5,-.87)--(.5,.87);
\draw [line,->-=.6] (-1,0) node [left] {\scriptsize $x$}
-- (-.5,0) node [above] {$c$} -- (0,0);
\draw [line,-<-=.56] (-1,0) -- (-1.5,.87) node [above left=-3pt] {$a$};
\draw [line,-<-=.56] (-1,0) -- (-1.5,-.87) node [below left=-3pt] {$b$};
\draw [line,-<-=.56] (0,0) -- (.5,-.87) node [below right=-3pt] {$m$};
\draw [line,->-=.56] (0,0) node [right] {\scriptsize $y$}
-- (.5,.87) node [above right=-3pt] {$n$};
\end{tikzpicture}
\end{gathered}
~~=~~
\sum_{k, z, w}
(\tF^{a, b, n}_{m; c, k})_{x, y}^{z, w}
\begin{gathered}
\begin{tikzpicture}[scale=1]
\fill[fill=gray!20] (.87,-1.5)--(0,-1)--(0,0)--(.87,.5);
\draw [line,->-=.56] (0,-1) node [below] {\scriptsize $z$}
-- (0,-.5) node [right] {$k$} -- (0,0);
\draw [line,-<-=.56] (0,-1) -- (.87,-1.5) node [below right=-3pt] {$m$};
\draw [line,-<-=.56] (0,-1) -- (-.87,-1.5) node [below left=-3pt] {$b$};
\draw [line,-<-=.56] (0,0) -- (-.87,.5) node [above left=-3pt] {$a$};
\draw [line,->-=.56] (0,0) node [above] {\scriptsize $w$}
-- (.87,.5) node [above right=-3pt] {$n$};
\end{tikzpicture}
\end{gathered}
\, .
\fe

The NIM-rep matrices $\tN$ and the F-symbols $\tF$ capture all the data of the category $\cM$ of BCs.
In the {categorical language}, $\cM$ is called a module category over $\cC$ or a $\cC$-module category.

\subsubsection*{Boundary OPE}

By the state-operator correspondence, the space of boundary-changing operators between BCs $m$ and $n$ is given by $\mathcal{H}_{m,n} := \mathcal{H}^\cI_{m,n}$.
One can fuse different boundary-changing operators and get an associative OPE algebra 
\ie
\begin{gathered}
\begin{tikzpicture}[scale=1]
\fill [fill=gray!20] (0,-1) rectangle (1,2);
\draw [line,->-=.55] (0,-1) -- (0,0);
\draw (0,-0.5) node [right] {$m$};
\draw [line,->-=.55] (0,0) -- (0,1);
\draw (0,0.5) node [right] {$n$};
\draw [line,->-=.55] (0,1) -- (0,2);
\draw (0,1.5) node [right] {$k$};
\filldraw[black] (0,0) circle (1pt);
\filldraw[black] (0,1) circle (1pt);
\end{tikzpicture}
\end{gathered} \quad \to \quad
\begin{gathered}
\begin{tikzpicture}[scale=1]
\fill [fill=gray!20] (0,-1.5) rectangle (1,1.5);
\draw [line,->-=.55] (0,-1.5) -- (0,0);
\draw (0,-.75) node [right] {$m$};
\draw [line,->-=.55] (0,0) -- (0,1.5);
\draw (0,.75) node [right] {$k$};
\filldraw[black] (0,0) circle (1pt);
\end{tikzpicture}
\end{gathered}
\, \quad:\quad
\mathcal{H}_{m,n} \times \mathcal{H}_{n,k} \to \mathcal{H}_{m,k}~.
\label{OPE}
\fe
In the {categorical language}, this OPE corresponds to the composition of morphisms in $\cM$ since $\mathcal{H}_{m,n} = \Hom_\cM(m,n)$.

We next introduce the disc partition functions (quantum dimensions) and disc one-point functions, which together form a trace structure on $\CM$.

\subsubsection*{Quantum dimensions}

We denote the quantum dimension of a BC $m$ by the value of its disc partition function, and denote it by $\td_m$,
\ie
\begin{gathered}
\begin{tikzpicture}[scale=0.7]
\begin{scope}[even odd rule]
\clip (0,0) circle (1)  (-2,-2) rectangle (2,2);
\fill [fill=gray!20] (-2,-2) rectangle (2,2);
\end{scope}
\draw [line,->-=.01,->-=.51] (0,0) circle (1);
\draw (1,0) node [right] {$m$};
\draw (-1,0) node [left] {$m$};
\end{tikzpicture}
\end{gathered}
= \td_m \, .
\fe
These quantum dimensions have to be compatible with the action of TDLs on BCs.
Namely, the partition function of a loop of TDL $a$ parallel to a disc boundary with BC $m$ can be evaluated in two ways:
we can first fuse $a$ with $m$ and evaluate the partition function or we can can first shrink $a$ and then evaluate the disc partition function,
\ie
& \sum_{n} \tN_{am}^n \, 
\begin{gathered}
\begin{tikzpicture}[scale=0.7]
\begin{scope}[even odd rule]
\clip (0,0) circle (1)  (-2,-2) rectangle (2,2);
\fill [fill=gray!20] (-2,-2) rectangle (2,2);
\end{scope}
\draw [line,->-=.01,->-=.51] (0,0) circle (1);
\draw (1,0) node [right] {$n$};
\draw (-1,0) node [left] {$n$};
\end{tikzpicture}
\end{gathered}=
\begin{gathered}
\begin{tikzpicture}[scale=0.7]
\begin{scope}[even odd rule]
\clip (0,0) circle (1)  (-2,-2) rectangle (2,2);
\fill [fill=gray!20] (-2,-2) rectangle (2,2);
\end{scope}
\draw [line,->-=.01,->-=.51] (0,0) circle (1);
\draw [line,->-=.02,->-=.52] (0,0) circle (.5);
\draw (1,0) node [right] {$m$};
\draw (-1,0) node [left] {$m$};
\draw (0,0) node {$a$};
\end{tikzpicture}
\end{gathered}=
\begin{gathered}
\begin{tikzpicture}[scale=0.7]
\begin{scope}[even odd rule]
\clip (0,0) circle (1)  (-2,-2) rectangle (2,2);
\fill [fill=gray!20] (-2,-2) rectangle (2,2);
\end{scope}
\draw [line,->-=.01,->-=.51] (0,0) circle (1);
\draw (1,0) node [right] {$m$};
\draw (-1,0) node [left] {$m$};
\draw [line] (0,0) node {$d_a$};
\end{tikzpicture}
\end{gathered}
\fe
Thus we find 
\ie
& \sum_{n} \tN_{am}^n \, \td_n = d_a \td_m \, ,
\label{BoundaryEigenvector}
\fe
which says that the boundary quantum dimensions $\td$ form a simultaneous eigenvector of the NIM-rep matrices $\tN_a$.
Unitarity/reflection-positivity requires the boundary quantum dimensions to be positive, and fixes $\td$ to be proportional to a Frobenius-Perron eigenvector of the NIM-rep matrices, which is unique for an indecomposable module category \cite{Etingof:aa}.
More generally, $\td$ of any indecomposable module category is unique up to an overall normalization \cite{schaumann2013traces}.
The ambiguous overall normalization physically reflects the fact that it can be rescaled by adjusting the coefficient of the Euler counter-term.
As we will see, this overall rescaling is also related to a rescaling of the trace in the bulk Frobenius algebra.

\subsubsection*{Disc one- and two-point functions}

More generally, we can compute the disc one-point function of boundary-changing operators, which gives a collection of linear maps
\ie
\label{trace}
\theta_m:\mathcal{H}_{m,m} \to \bC \, ,
\qquad
x \mapsto
\begin{gathered}
\begin{tikzpicture}[scale=0.7]
\begin{scope}[even odd rule]
\clip (0,0) circle (1)  (-2,-2) rectangle (2,2);
\fill [fill=gray!20] (-2,-2) rectangle (2,2);
\end{scope}
\draw [line,->-=.01,->-=.51] (0,0) circle (1);
\draw (1,0) node [right] {$m$};
\draw (-1,0) node [left] {$m$};
\filldraw[black] (0,-1) circle (1.5pt) node[below] {\scriptsize $x$};
\end{tikzpicture}
\end{gathered}
\, .
\fe
In a unitary/reflection-positive, or more generally semisimple TFT, the Hilbert space $\mathcal{H}_{m,m}$ is one-dimensional for a simple BC $m$.
Therefore, these linear maps are essentially determined by the boundary quantum dimensions $\td_m$.
Combining these linear maps with the OPE structure in \eqref{OPE}, we get the disc two-point function of boundary-changing operators.

The collection of linear maps in \eqref{trace} is called a trace on the category $\cM$.\footnote{Categories equipped with a trace are called Calabi-Yau categories in the literature \cite{costello2007topological}.
} When the trace is furthermore compatible with the $\cC$-module structure, i.e.\ satisfying equation \eqref{BoundaryEigenvector}, it is called a \emph{module trace}~\cite{schaumann2013traces}.
The trace defines a symmetric non-degenerate pairing between Hom spaces
\begin{equation}
    \label{TracePairing}
    \mathrm{Hom}_\cM(m,n) \times \mathrm{Hom}_\cM(n,m) \to \bC\,, \qquad (f,g) \mapsto \theta_m(g\circ f) = \theta_n(f\circ g) \, .
\end{equation}
As was argued in \cite{Moore:2006dw}, a unitary TFT (without $\cC$-structure) is completely determined by the category $\cM$ of BCs and the trace.
Thus, forgetting the symmetry $\cC$, the category $\cM$ is enough to determine the observables in the TFT.
Taking into account the TDLs in $\cC$, we can ask if knowing the $\cC$-module category $\cM$ is enough to construct all the observables of the $\cC$-symmetric TFT, including defect correlators.
The answer is yes, as we will show in section \ref{Sec:Axiomatic}.

A category $\cM$ of BCs that always exists for any $\cC$ is one isomorphic to $\cC$ itself: we can take the NIM-reps $\tN$ and the F-symbols $\tF$ of $\cM$ to be the same as the fusion coefficients $N$ and the F-symbols $F$ of $\cC$.
Formally, this corresponds to viewing $\cC$ as a module category over itself, which is known as the \emph{regular} module category.
In this case, the $\td$ of $\cM$ coincide with the $d$ in $\cC$, and defines a trace structure on $\cM$.

\subsubsection*{Adjoint structure and orientation-reversal}

We can insert a TDL in the middle of a disc and compute the two-point function of boundary-changing defect operators
\begin{equation}
    \mathcal{H}^a_{m,n} \times \mathcal{H}^{\bar{a}}_{n,m} \to \bC \,.
\end{equation}
Similar to the TDL case, one can define an antilinear adjoint map 
\ie
\dag : \quad \mathcal{H}^a_{m,n} \to \mathcal{H}^{\bar{a}}_{n,m} \, ,
\fe
such that the disc two-point functions defines a positive-definite
bilinear form on the junction vector spaces
\ie \label{inner.product}
\vev{y^\dagger x}  \quad := \quad \Thjfill{n}{a}{m}{y^\dagger}{x}~,
\fe
where $x,y\in \mathcal{H}^a_{m,n}$.
The relation between $x$ and $x^\dag$ can again be understood physically as an orientation-reversal on the junction.
In a non-unitary TFT, this bilinear form is not positive-definite but is still required to be non-degenerate.

\section{Construction of topological field theories with non-invertible symmetries}
\label{Sec:Axiomatic}

The data of a TFT (without $\cC$ structure) consists of the Hilbert space of bulk local operators, the Hilbert spaces of boundary-changing operators, together with their correlators which can generally be built by cutting-and-sewing from a set of {defining data}.
When there is a fusion category symmetry $\cC$ acting on the theory, the TFT data also includes the correlators in the presence of TDLs in $\cC$.
Therefore, the data of a $\cC$-symmetric TFT consists of open and closed defect Hilbert spaces and their general correlation functions, which can also be built from the {defining data}.

Given a symmetry $\cC$ and a $\cC$-module category $\cM$ describing all possible BCs, we now explicitly construct the {defining data} of a TFT.

\subsection{Open and closed Hilbert spaces}
\label{Sec:Hilbert}

The open defect Hilbert space $\cH^a_{m,n}$ is a vector space of point-like boundary-changing defect operators at the junction of the bulk TDL $a$ with BCs $m$ and $n$
\ie
\label{BoundaryChangingOperator}
\begin{gathered}
\begin{tikzpicture}
\begin{scope}[even odd rule]
\clip (0,0) ++(180:1) arc (180:360:1)  (-1.2,-1.2) rectangle (1.2,0);
\fill [fill=gray!20] (-1.2,-1.2) rectangle (1.2,0);
\end{scope}
\draw [line,->-=.25,->-=.8] (0,0) ++(180:1) arc (180:360:1);
\draw [line,->-=.6] (0,-1) node[below] {} -- (0,-.5) node[right] {$a$} -- (0,0);
\draw (-.72,-.72) node[below left] {$m$};
\draw (.72,-.72) node[below right] {$n$};
\draw [fill=black] (0,-1) circle (.05);
\end{tikzpicture}
\end{gathered} \, .
\fe
Under the operator-state map, this vector space is the same as the defect Hilbert space of the TFT on an interval, with BCs $m$ and $n$ and twisted by the TDL $a$ in between,
\ie
\begin{gathered}
\begin{tikzpicture}[scale=1]
\fill [fill=gray!20] (0,-1) rectangle (1,1);
\draw [line,->-=.55] (0,-1) -- (0,0);
\draw (0,-0.6) node [right] {$m$};
\draw [line,->-=.55] (0,0) -- (0,1);
\draw (0,0.6) node [right] {$n$};
\draw [line,->-=.55] (0,0) -- (-1,0);
\draw (-0.6,0) node [below] {$a$};
\filldraw[black] (0,0) circle (1pt) node[right] {\scriptsize $x$};
\end{tikzpicture}
\end{gathered} 
\qquad
\leftrightarrow
\qquad
\begin{gathered}
\begin{tikzpicture}[scale=1, ,rotate=-90]
\fill [fill=gray!20] (-1,.5) rectangle (1,1);
\fill [fill=gray!20] (-1,-1) rectangle (1,-.5);
\draw [line,->-=.55] (1,.5) -- (-1,.5);
\draw (-.2,.5) node [right] {$n$};
\draw [line,->-=.55] (1,0) -- (-1,0);
\draw (-.2,0) node [left] {$a$};
\filldraw[black] (1,0) circle (1pt) node[below] {$\cB_{m,n}^{a;x}$};
\filldraw[black] (-1,0) node[above] {\phantom{$\cB_{m,n}^{a;x}$}};
\draw [line,-<-=.55] (1,-.5) -- (-1,-.5);
\draw (-.2,-.5) node [left] {$m$};
\end{tikzpicture}
\end{gathered} \, .
\fe
As was discussed in the previous section, this is identified with $\Hom_\cM(m, a \otimes n)$ in the {categorical language}.
In other words, for any junction vector $x \in \Hom_\cM(m, a \otimes n)$, there is a boundary-changing defect operator that we denote by
\begin{equation}
    \BCO^{a;x}_{m,n} \in \cH^a_{m,n} \, .
    \label{open.hilbert.space}
\end{equation}

The closed defect Hilbert space $\cH^a$ is the vector space of point-like bulk (as opposed to boundary-changing) defect operators at the end of a TDL $a$.
This space cannot be immediately identified with a junction vector space in $\cC$ or in $\cM$.
Instead, a closed Hilbert space is to be constructed from an open Hilbert spaces via the open-to-closed map, which is the following.
First, we remove a disc around some point in the bulk and impose the BC $m$ on the boundary of the disc.
Then we attach the TDL $a$ to the BC $m$ via the boundary-changing defect operator $\BCO^{a;x}_{m,m}$.
Finally, shrinking the boundary disc to a point results in a bulk defect operator, which we denote by $\cO^{a;x}_m \in \cH^a$,
\ie
\begin{gathered}
\begin{tikzpicture}[scale=.5,rotate=270]
\draw [line, -<-=0, fill=gray!20] (0,0) circle (.6);
\draw [fill=black] (-0.6,0) circle (.1);
\draw [line, -<-=.5] (-3,0) -- (-0.6,0);
\draw (-3,0) node [left] {$a$};
\draw (-1,0) node [right] {\scriptsize $x$};
\draw (0.6,0) node [below] {$m$};
\end{tikzpicture}
\end{gathered}\quad\sim\quad
\begin{gathered}
\begin{tikzpicture}[scale=.5,rotate=270]
\draw [fill=black] (0.6,0) circle (.1);
\draw [line, -<-=.5] (-3,0) -- (0.6,0);
\draw (-3,0) node [left] {$a$};
\draw (0.6,0) node [right] {$\cO^{a;x}_m$};
\end{tikzpicture}
\end{gathered}~.
\label{bc.op}
\fe
For any TDL $a$, this construction gives the open-to-closed linear map
\ie
\bigoplus_m \cH^a_{m,m} \to \cH^a \, , \qquad \BCO^{a;x}_{m,m} \mapsto \cO^{a;x}_m \, , \label{boundary.bulk}
\fe
which we show to be an isomorphism below.

For the Hilbert space of local operators, i.e.\ when $a=\cI$, the isomorphism property has been proven in \cite{Moore:2006dw} from the definition that $\cM$ is the category of \emph{all} BCs, or in other words, the \emph{maximal} category of BCs.
We will borrow their result to prove the isomorphism property for $a\neq \cI$.
In particular, we construct a closed-to-open map from $\cH^a$ to $\bigoplus_m \cH^a_{m,m}$ for any $a$, and show that the open-to-closed and closed-to-open maps are both injective.
Note that because they are maps between finite-dimensional vector space, two-sided injectivity already implies that the open-to-closed and closed-to-open maps are isomorphisms.
The injectivity of the open-to-closed maps will be shown in section \ref{Sec:BoundaryCrossing} to follow from boundary crossing relations, while the injectivity of the closed-to-open maps is shown below.

A closed-to-open map is defined by taking a bulk defect operator in $\cH^a$ and pushing it to a BC $m$, to obtain a defect operator in $\cH^a_{m,m}$.
Assume that there is a nonzero defect operator $\cO \in \cH^a$ in the kernel of the closed-to-open map.
Since $\cO$ is nonzero, there exists another defect operator ${\ocO} \in \cH^{\bar{a}}$ such that the local operator
\begin{equation}
    \cO {\ocO} = 
    \begin{gathered}
    \begin{tikzpicture}
    \draw [fill=black] (0,0) circle (.05) node [above] {$\cO$};
    \draw [fill=black] (1,0) circle (.05) node [above] {${\ocO}$};
    \draw [line, ->-=.6] (0,0) -- (1,0);
    \draw (0.5,0) node [below] {$a$};
    \end{tikzpicture}
    \end{gathered}
    \in \cH
\end{equation}
is nonzero.
Since the closed-to-open map is injective for all local operators ($a = \cI$) by \cite[theorem~2]{Moore:2006dw}, there must exist some BC $m$ such that when we push the local operator $\cO \ocO$ to the boundary, we get a nonzero boundary-changing operator.
But 
we can alternatively first push $\cO$ towards the BC $m$
to get zero by the initial assumption that $\cO$ is in the kernel of the closed-to-open map.
A contradiction is thus reached, showing that $\cO$ cannot exist.
Therefore, we have proven that the closed-to-open map is injective for any $a$.

All in all, we get the isomorphism $\cH^a \cong \bigoplus_{m} \Hom_\cM(m, a \otimes m)$, which in particular implies that
\ie
\label{dimH}
\text{dim} \, \cH^a = \sum_{m} \tN_{am}^m \, .
\fe
After introducing the action of TDLs on bulk local operators, we will see that this is consistent with torus modular invariance.
In the following we omit the superscript $\cI$ for the Hilbert space of local operators, and set $\cH := \cH^\cI$.

\subsection{Open correlators}
\label{Sec:Open}

Correlators of purely boundary-changing operators can be computed from the module category $\cM$ and its trace in the following way.

First we compute disc correlators.
On the boundary of the disc, pick an arbitrary segment with BC $m$, and regard the correlator as a trace over a morphism in $\Hom_\cM(m, m)$.
Then use the OPE of boundary operators, i.e.\ the composition of morphisms in $\cM$, to produce a single operator on the boundary.
This reduces the correlator to a disc one-point function, which is nothing but the module trace \eqref{trace}.
Hence disc correlators are completely captured by the module category $\cM$ and its trace.

Open correlators on more general topologies can be decomposed via cutting-and-sewing into disc correlators and closed correlators.
Given that we have already constructed the disc correlators, the construction of general open correlators amounts to the construction of general closed correlators, which we subsequently undertake.
But before that, we need to introduce a key ingredient, the boundary crossing relation.

\subsection{Boundary crossing relation}
\label{Sec:BoundaryCrossing}

For two simple BCs $m$ and $n$, the open-sector Hilbert space $\cH^\cI_{m,n}$ is one-dimensional if $m = n$, and empty if $m \neq n$.
Hence, the cutting-and-sewing of a strip bounded by two simple BCs implies that
\ie
\label{BoundaryCrossing}
\begin{gathered}
\begin{tikzpicture}[scale=1]
\fill [fill=gray!20] (-1.4,0) ++(45:1) arc (45:-45:1);
\fill [fill=gray!20] (1.4,0) ++(225:1) arc (225:135:1);
\draw [line,->-=.55] (-1.4,0) ++(45:1) arc (45:-45:1);
\draw [line,->-=.55] (1.4,0) ++(225:1) arc (225:135:1);
\draw (-.5,0) node [left] {$n$};
\draw (.5,0) node [right] {$m$};
\end{tikzpicture}
\end{gathered}
~~ = ~~
\frac{\D_{m,n}}{\td_m}
\quad
\begin{gathered}
\begin{tikzpicture}[xscale=1,rotate=90]
\begin{scope}[even odd rule]
\clip (-0.693,-0.707) rectangle (0.693,0.707) (-1.4,0) ++(45:1) arc (45:-45:1) (1.4,0) ++(225:1) arc (225:135:1);
\fill [fill=gray!20] (-0.693,-0.707) rectangle (0.693,0.707);
\end{scope}
\draw [line,-<-=.55] (-1.4,0) ++(45:1) arc (45:-45:1);
\draw [line,-<-=.55] (1.4,0) ++(225:1) arc (225:135:1);
\draw (-.5,0) node [below] {$m$};
\draw (.5,0) node [above] {$m$};
\end{tikzpicture}
\end{gathered}
\, ,
\fe
which we call the {\it boundary crossing relation}.
Since the boundaries are topological, this relation can also be interpreted as an F-move in a larger categorical structure that contains $\cC$ and $\cM$.

Let us interpret the boundary crossing relation in module category terms.
Suppose the lefthand side of \eqref{BoundaryCrossing} is the local configuration of a general disc 
\ie
\label{Blobs}
\begin{gathered}
\begin{tikzpicture}[scale=.75]
\begin{scope}[even odd rule]
\clip (0,0) circle (1)  (-1.6,-1.6) rectangle (1.6,1.6);
\fill [fill=gray!20] (-1.6,-1.6) rectangle (1.6,1.6);
\end{scope}
\draw [line,->-=.01,->-=.51] (0,0) circle (1);
\draw [fill=black] (0,1) circle (.25);
\draw [fill=black] (0,-1) circle (.25);
\draw (-1,0) node [left] {$n$};
\draw (1,0) node [right] {$m$};
\end{tikzpicture}
\end{gathered} \, ,
\fe
where each blob represents a general boundary with an arbitrary configuration of TDLs attached, and importantly, that the two blobs have no TDL stretched between.
By regarding the above as the module trace of
\ie
\begin{gathered}
\begin{tikzpicture}[scale=.75]
\fill [fill=gray!20] (0,-3) rectangle (2,3);
\draw [line,->-=.17,->-=.5,->-=.85] (0,-3) -- (0,-2) node [right] {$m$} -- (0,0) node [right] {$n$} -- (0,2) node [right] {$m$} -- (0,3);
\draw [fill=black] (0,1) node [right=7pt] {\scriptsize $y$} circle (.25);
\draw [fill=black] (0,-1) node [right=7pt] {\scriptsize $x$} circle (.25);
\end{tikzpicture}
\end{gathered}
\fe
where $x \in \Hom_\cM(m,n)$ and $y \in \Hom_\cM(n,m)$ are morphisms corresponding to the blobs, it is immediately clear that if $m$ and $n$ are simple, then it is zero unless $n = m$.
Moreover, because $\Hom_\cM(m,m)$ is one-dimensional, we can write $x = X 1_m$ and $y = Y 1_m$ where $X, Y \in \mathbb{C}^\times$.
Then \eqref{Blobs} evaluates to
\ie
\label{XY}
X Y \td_m \, .
\fe
On the other hand, performing \eqref{BoundaryCrossing} on \eqref{Blobs} and then evalating the module traces gives
\ie
\frac{1}{\td_m} (X \td_m) \, (Y \td_m) \, ,
\fe
which is obviously equal to \eqref{XY}.
Since the blobs are general, we have proven \eqref{BoundaryCrossing}.

We can prove a stronger boundary crossing relation.
Consider the configuration on the lefthand side of \eqref{BoundaryCrossing}, but with an extra TDL $a$ parallel to the boundaries.
In this situation, we can interpret the lefthand side as the identity map from the Hilbert space $\cH_{m,n}^a$ to itself.
But we can insert a complete set of basis states on a horizontal cut, and rewrite this identity map as
\begin{equation}
    \sum_{x,y} \ket{\BCO_{m,n}^{a;x}} \, G_{xy} \, \bra{\BCO_{n,m}^{\bar{a};y}}~,
\end{equation}
where $x$ and $y$ in the summation run over complete bases of $\cH_{m,n}^a$ and $\cH_{n,m}^{\bar{a}}$, respectively, and the matrix $G$ is the inverse disc two-point function
\begin{equation} \label{g.inverse}
    \left( G^{-1} \right)^{yx} = {\left \langle \BCO_{n,m}^{\bar{a};y} \,  \BCO_{m,n}^{a;x} \right \rangle}_{D^2} = \Thjfill{n}{a}{m}{y}{x}~.
\end{equation}
Thus, we have proven
\ie
\label{GeneralizedBoundaryCrossing}
\begin{gathered}
\begin{tikzpicture}[scale=1]
\draw [line,->-=.55, fill=gray!20] (-1.4,0) ++(45:1) arc (45:-45:1);
\draw [line,->-=.55, fill=gray!20] (1.4,0) ++(225:1) arc (225:135:1);
\draw [line,->-=.55] (0,-0.75) node[below] {} -- (0,0.5) node[right] {$a$} -- (0,0.75);
\draw (-.5,0) node [left] {$m$};
\draw (.5,0) node [right] {$n$};
\end{tikzpicture}
\end{gathered}
~~ = ~~
\sum_{x,y} G_{xy}
~
\begin{gathered}
\begin{tikzpicture}[scale=0.85]
\begin{scope}[even odd rule]
\def\arca{(0,0) ++(180:1) arc (180:360:1)}
\def\arcb{(0,-2.5) ++(180:1) arc (180:0:1)}
\clip (-1.7,-2.5) rectangle (1.7, 0) \arca \arcb;
\fill [fill=gray!20] (-1.7,-2.5) rectangle (1.7, 0);
\end{scope}
\draw [line,->-=.25,->-=.8] (0,0) ++(180:1) arc (180:360:1);
\draw [line,->-=.6] (0,-1) node[below] {} -- (0,-.3) node[right] {$a$} -- (0,0);
\draw (0,-1) node[above left] {\scriptsize $x$};
\draw (-1,0) node[below left] {$m$};
\draw (1,0) node[below right] {$n$};
\draw [line,-<-=.25,-<-=.8] (0,-2.5) ++(180:1) arc (180:0:1);
\draw [line,-<-=.6] (0,-1.5) node[below] {} -- (0,-2.2) node[left] {$a$} -- (0,-2.5);
\draw (0,-1.5) node[below right] {\scriptsize  $y$};
\draw (-1,-2.5) node[above left] {$m$};
\draw (1,-2.5) node[above right] {$n$};
\end{tikzpicture}
\end{gathered}
\, .
\fe

As promised earlier in this section, we now use the boundary crossing relation to prove that the open-to-closed map \eqref{boundary.bulk} is injective.
Assume the contrary, that there exists a boundary-changing defect operator $\sum_m \BCO^{a;x_m}_{m,m} \in \bigoplus_m \cH^a_{m,m}$ in the kernel for some junction vectors $x_m$.
Then the bulk defect operator $\sum_m \cO^{a;x_m}_{m} \in \cH^a$ vanishes, and hence any two-point function involving it must vanish as well.\footnote{The correlator $\vev{\cO_1 \cO_2 \cdots}$ without any subscript always stands for the (bulk) sphere correlator, i.e.\ $\vev{\cO_1 \cO_2 \cdots}=\vev{\cO_1 \cO_2 \cdots}_{S^2}$.
}
Thus for any 
$\cO^{\bar{a};y}_{n} \in \cH^{\bar{a}}$ 
we get
\begin{equation}
    \vev{ \left( \sum_m \cO^{a;x_m}_{m} \right) \cO^{\bar{a};y}_{n} } = 0~.
\end{equation}
Using the boundary crossing relation \eqref{BoundaryCrossing}, we can rewrite this correlator as a disc two-point function of boundary operators,
\ie
\sum\limits_{m}
\begin{gathered}
\begin{tikzpicture}[scale=.5]
\draw [line, -<-=0, fill=gray!20] (-3.6,0) circle (.6);
\draw [line, -<-=0, fill=gray!20] (0,0) circle (.6);
\draw [fill=black] (-0.6,0) circle (.1);
\draw [fill=black] (-3,0) circle (.1);
\draw [line, ->-=.5] (-3,0) -- (-0.6,0);
\draw (-1.5,0) node [below] {$a$};
\draw (-2.6,0) node [above] {\scriptsize $x_m$};
\draw (-1,0) node [above] {\scriptsize $y$};
\draw (0.6,0) node [right] {$n$};
\draw (-4.2,0) node [left] {$m$};
\end{tikzpicture}
\end{gathered}\to
\sum\limits_{m} \frac{\delta_{m,n}}{\td_n}
\begin{gathered}
\begin{tikzpicture}[scale=0.6]
\begin{scope}[even odd rule]
\def\arca{(0,1.43) ++(225:1) arc (225:315:1)}
\def\arcb{(0,-1.43) ++(135:1) arc (135:45:1);}
\clip (-0.71,-0.71) rectangle (0.71,0.71) \arca \arcb;
\fill [fill=gray!20] (-0.71,-0.71) rectangle (0.71,0.71);
\end{scope}
\draw [line, fill=gray!20] (-1.4,0) ++(45:1) arc (45:315:1);
\draw [line,->-=.57] (0,1.43) ++(225:1) arc (225:315:1);
\draw [line,-<-=.53] (0,-1.43) ++(135:1) arc (135:45:1);
\draw [line, fill=gray!20] (1.4,0) ++(135:1) arc (135:-135:1);
\draw (0,0.6) node [above] {$n$};
\draw [line, ->-=.5] (-0.7,-0.7) -- (0.7,-0.7);
\draw [fill=black] (-0.7,-0.7) circle (.1);
\draw [fill=black] (0.7,-0.7) circle (.1);
\draw (-0.9,-0.7) node [above] {\scriptsize $x_m$};
\draw (0.9,-0.7) node [above] {\scriptsize $y$};
\draw (0,-0.7) node [below] {$a$};
\end{tikzpicture}
\end{gathered}\to
\frac{1}{\td_n}\Thjfill{n}{a}{n}{y}{x_n} \, ,
\fe
and conclude that
\begin{equation}
    \frac{1}{\td_n} {\left \langle  {\cO^{a;x_n}_{n,n}}  \, \cO^{\bar{a};y}_{n,n} \right \rangle}_{D^2} = 0~.
\end{equation}
But this violates the non-degeneracy of the two-point function (see near \eqref{inner.product}) unless $x_n = 0$ for all BCs $n$.
Therefore, the open-to-closed map is injective.

\subsection{Closed correlators without defects}
\label{Sec:CWD}

The correlators of bulk local operators in a TFT is famously captured by a commutative Frobenius algebra \cite{Sawin:1995rh,Abrams:1996ty}.
Here we review how it is derived from the boundary data $\cM$.
The $\cC$-structure is not crucial here and will be ignored.

\subsubsection*{Frobenius Algebras}
A commutative Frobenius algebra is a commutative associative algebra that is equipped with a trace.
In the TFT context, the commutative associative algebra structure is given by the OPE of bulk operators, and the trace is given by the $S^2$ one-point function.
In a unitary/reflection-positive TFT, the algebra is moreover semisimple.
Such semisimple algebras decompose into decoupled copies of one-dimensional algebras.
In other words, a unitary TFT without any defect decomposes into decoupled TFTs, which we label by $i$, with one-dimensional Hilbert spaces $\cH_i = \{ \pi_i \}$ \cite{Durhuus:1993cq,Sawin:1995rh}.
By decoupled we mean that $\pi_i$ are mutually orthogonal, and since each Hilbert space is one-dimensional, $\pi_i^2$ is proportional to $\pi_i$ ($\pi_i^2 = 0$ violates semisimplicity).
It is conventional to normalize $\pi_i$ into a complete set of idempotents (orthogonal projectors) satisfying
\begin{equation}
    \pi_i \pi_j = \delta_{ij} \pi_i~.
\end{equation}
These idempotents form a basis of the Hilbert space $\cH$ on the circle.
Each one-dimensional TFT is almost trivial, with the trace/one-point function $\pi_i \mapsto \langle \pi_i \rangle = \lambda_i$ being the only parameter, whose logarithm times the Euler number gives (twice) the action.
Putting the decoupled copies together, the partition function of the full TFT on a genus $g$ surface is given by
\begin{equation}
    Z[\Sigma_g] = \sum_i \lambda_i^{1-g}~.
\end{equation}
In particular, we have $1=\sum_i \pi_i$ and therefore $Z[S^2] = \langle 1 \rangle = \sum_i \lambda_i$.
A unitary TFT further satisfies $\lambda_i > 0$.

\subsubsection*{Idempotents from Boundary Conditions}

Now we use the open-to-closed map to construct the Frobenius algebra from boundaries.
The map gives the isomorphism
\begin{equation}
     \cH \simeq \bigoplus_m \cH_{m,m} \, .
\end{equation}
Each $\cH_{m,m}$ is one-dimensional,
hence for each simple BC $m$, we get a basis operator
    $\cO_m \in \cH~$
created by a circular BC $m$.\footnote{In the notation we used around \eqref{bc.op}, $\cO_m := \cO_m^{\mathbbm{1};1_m}$ where $1_m \in \mathrm{Hom}_\cM(m,m)$.}

To obtain the Frobenius algebra structure, we have to compute the OPE of basis operators $\cO_m$.
We insert two such operators near each other, which corresponds to putting BCs $m$ and $n$ on the boundary of two nearby empty discs.
From the boundary crossing relation \eqref{BoundaryCrossing}, we have
\ie
\label{MergeGraph}
\begin{gathered}
\begin{tikzpicture}[scale=1]
\fill [fill=gray!20] (-1.4,0) circle (1);
\fill [fill=gray!20] (1.4,0) circle (1);
\draw [line] (-1.4,0) ++(45:1) arc (45:315:1);
\draw [line,->-=.55] (-1.4,0) ++(45:1) arc (45:-45:1);
\draw [line,->-=.55] (1.4,0) ++(225:1) arc (225:135:1);
\draw [line] (1.4,0) ++(135:1) arc (135:-135:1);
\draw (-.5,0) node [left] {$n$};
\draw (.5,0) node [right] {$m$};
\end{tikzpicture}
\end{gathered}\quad = \quad\frac{\D_{m,n}}{\td_m}
\begin{gathered}
\begin{tikzpicture}[scale=1]
\begin{scope}[even odd rule]
\def\arca{(0,1.43) ++(225:1) arc (225:315:1)}
\def\arcb{(0,-1.43) ++(135:1) arc (135:45:1);}
\clip (-0.7,-0.7) rectangle (0.7,0.7) \arca \arcb;
\fill [fill=gray!20] (-0.7,-0.7) rectangle (0.7,0.7);
\end{scope}
\draw [line, fill=gray!20] (-1.4,0) ++(45:1) arc (45:315:1);
\draw [line,->-=.57] (0,1.43) ++(225:1) arc (225:315:1);
\draw [line,-<-=.53] (0,-1.43) ++(135:1) arc (135:45:1);
\draw [line, fill=gray!20] (1.4,0) ++(135:1) arc (135:-135:1);
\draw (0,0.6) node [above] {$m$};
\end{tikzpicture}
\end{gathered} \, .
\fe
Using the module trace, we deduce that the Frobenius algebra is given by
\ie
\label{Frobenius}
\cO_m \cO_n = \frac{\D_{m,n}}{\td_m} \cO_m \, , \qquad \vev{\cO_m} = \td_m \, .
\fe
We see that the one-point function, which is called the trace of the bulk Frobenius algebra, is rescaled accordingly when the overall normalization of $\td$ is rescaled.
As mentioned before, this overall normalization is unphysical because it can be changed by adjusting the coefficient of the Euler counter-term.

The local operators $\cO_m$ form an orthonormal basis, i.e.\ $\langle \cO_m \cO_n \rangle = \delta_{mn}$, which is sometimes called the Cardy basis in analogy with boundary states in conformal field theories.
With a slight change of normalization, we can relate the Cardy basis to the idempotents (orthogonal projectors) introduced earlier,
\ie
\left\{ \pi_m \equiv \td_m \cO_m \right\} \, , \qquad \vev{\pi_m} = \td_m^2 \, .
\fe
The above construction of the Frobenius algebra of a $\cC$-symmetric TFT was originally given in~\cite[appendix A.3]{Komargodski:2020mxz}.
The partition function of the TFT is given by
\ie
\label{IdentityNormalization}
Z(\Sigma_g) = \sum_m \td_{m}^{2(1-g)} \, .
\fe

\subsubsection*{Action of TDLs on Frobenius algebra}
It follows from the fusion \eqref{BoundaryFusion} of parallel TDLs and BCs that
\ie
\widehat a \cdot \cO_m = \sum_{n} \tN_{am}^n \cO_n \, ,
\fe
where $\widehat a$ denotes an operator on the bulk Frobenius algebar representing the encircling and shrinking of a closed TDL $a$ on $\cO \in \cH$, 
\ie
\begin{gathered}
\begin{tikzpicture}[scale=.5]
\draw [line,->-=0] (0,0) circle (1.5);
\draw [fill=black] (0,0) circle (0.1) node[above] {$\cO$};
\draw (-1.5,0) node [left] {$a$};
\end{tikzpicture}
\end{gathered}\quad = \quad
\begin{gathered}
\begin{tikzpicture}[scale=.5]
\draw [fill=black] (0,0) circle (0.1);
\draw (0,0) node[above] {$\widehat a \cdot \cO$};
\end{tikzpicture}
\end{gathered}.
\fe
In other words, $\cH$ is a module over the fusion algebra of TDLs in $\cC$.
The torus partition function dressed with a TDL $a$ supported on a spatial cycle
\ie
Z\left(~
\begin{gathered}
\begin{tikzpicture}[scale = .5]
\draw [bg] (0,0) -- (2,0) -- (2,2) -- (0,2) -- (0,0);
\draw [line, ->-=0.6] (0,1) -- (1,1) node[above] {$a$} -- (2,1);
\end{tikzpicture}
\end{gathered}
~\right)
 = \text{Tr}_{\cH}(\wh{a})
\fe
equals $\sum_m \tN_{am}^m$.
Comparing with $\eqref{dimH}$, interpreted as the torus partition function with $a$ along the temporal circle
\ie
Z\left(~\begin{gathered}
\begin{tikzpicture}[scale = .5]
\draw [bg] (0,0) -- (2,0) -- (2,2) -- (0,2) -- (0,0);
\draw [line, ->-=0.55] (1,0) -- (1,1) node[right] {$a$} -- (1,2);
\end{tikzpicture}
\end{gathered}
~\right) = \text{Tr}_{\cH^a}(1) \, ,
\fe
we see that torus modular invariance is always satisfied!

\subsection{Closed correlators with defects}
\label{Sec:Crossing}

First, we derive general correlators on the sphere.
Correlators on general surfaces are then obtained by cutting-and-sewing.
Moreover, we rigorously prove that our formula for the three-point function solves crossing symmetry.\footnote{As for the torus one-point modular invariance, which is also needed to show that the general correlator is independent of how the cutting-and-sewing is done, we presently do not have a mathematically rigorous proof.
However, our formula is a necessary consequence of the full consistency of the TFT.
See \cite{Thorngren:2019iar} for an alternative physical argument for why the full consistency is expected.
}

\subsubsection*{Correlators on the sphere}
In the closed Hilbert space $\cH^a$ of point-like defect operators, we have learned that $\cO^{a;x}_m$ is created by a circular BC $m$ that is twisted by $a$ with junction vector $x$.
A general closed-sector correlator on the sphere can be computed by the following prescription.
\begin{enumerate}
\item  Merge all boundaries into one by the use of F-moves \eqref{F} and \eqref{BoundaryF} and the boundary crossing relation \eqref{GeneralizedBoundaryCrossing}.
\begin{enumerate}

\item[1a.] For any two disconnected boundaries, after using isotopy and the F-move \eqref{F} in $\cC$, we can assume that on a certain segment of the strip, two boundaries are separated by a single TDL.
We illustrate this procedure for initial separation by two TDLs $a$ and $b$:
\ie
\begin{gathered}
\begin{tikzpicture}[scale=.5]
\fill [fill=gray!20] (-3.5,0) circle (1.5);
\fill [fill=gray!20] (3.5,0) circle (1.5);
\draw [line,-<-=.05,-<-=.55] (-3.5,0) circle (1.5);
\draw [line,-<-=.05,-<-=.55] (3.5,0) circle (1.5);
\draw [line,-<-=.5] (-1,-2) -- (-1,2);
\draw [line,->-=.5] (1,-2) -- (1,2);
\draw (-1,0) node [right] {$a$};
\draw (1,0) node [left] {$b$};
\end{tikzpicture}
\end{gathered}\quad \rightarrow\quad
\begin{gathered}
\begin{tikzpicture}[scale=.5]
\fill [fill=gray!20] (-3.5,0) circle (1.5);
\fill [fill=gray!20] (3.5,0) circle (1.5);
\draw [line,-<-=.05,-<-=.55] (-3.5,0) circle (1.5);
\draw [line,-<-=.05,-<-=.55] (3.5,0) circle (1.5);
\draw [line,->-=.25,->-=.85] (0,2) ++(180:1) arc (180:360:1);
\draw [line,->-=.25,->-=.85] (0,-2) ++(0:1) arc (0:180:1);
\draw [line,->-=.5] (0,-1) -- (0,1);
\draw (-1,2) node [left] {$a$};
\draw (-1,-2) node [left] {$a$};
\draw (1,2) node [right] {$b$};
\draw (1,-2) node [right] {$b$};
\draw (0,0) node [right] {$c$};
\end{tikzpicture}
\end{gathered}
\fe

\item[1b.]  Merge two boundaries into one by use of the boundary crossing relation \eqref{GeneralizedBoundaryCrossing}.
\ie
\begin{gathered}
\begin{tikzpicture}[scale=.5]
\fill [fill=gray!20] (-2.1,0) circle (1.5);
\fill [fill=gray!20] (2.1,0) circle (1.5);
\draw [line,->-=.05,->-=.55] (-2.1,0) circle (1.5);
\draw [line,->-=.05,->-=.55] (2.1,0) circle (1.5);
\draw [line,->-=.5] (0,-2) -- (0,2);
\draw (0,-2) node [right] {$c$};
\end{tikzpicture}
\end{gathered}\quad \rightarrow\quad
\begin{gathered}
\begin{tikzpicture}[scale=.5]
\begin{scope}[even odd rule]
\def\arca{(0,2.12) ++(225:1.5) arc (225:315:1.5)}
\def\arcb{(0,-2.12) ++(135:1.5) arc (135:45:1.5);}
\clip (-1.062,-1.062) rectangle (1.062,1.062) \arca \arcb;
\fill [fill=gray!20] (-1.062,-1.062) rectangle (1.062,1.062);
\end{scope}
\draw [line, fill=gray!20, ->-=0.5] (-2.12,0) ++(45:1.5) arc (45:315:1.5);
\draw [line] (0,2.12) ++(225:1.5) arc (225:315:1.5);
\draw [line] (0,-2.12) ++(135:1.5) arc (135:45:1.5);
\draw [line, fill=gray!20, -<-=0.45] (2.12,0) ++(135:1.5) arc (135:-135:1.5);
\draw [line,->-=.5] (0,0.62) -- (0,2) node[right]{$c$};
\draw [line,->-=.5] (0,-2)node[right]{$c$} -- (0,-0.62);
\end{tikzpicture}
\end{gathered}
\fe

\item[1c.]  Repeat 1a and 1b until all disconnected boundaries are merged.

\end{enumerate}

\item Regard the sphere with one boundary as a disc with a certain BC and dressed with a network of TDLs, and evaluate it using F-moves and the module trace.
\end{enumerate}

We now use this prescription to compute up to four-point correlators on the sphere, and prove that the three-point function satisfies crossing symmetry.

Consider (the TDL types are implicit in the superscripts of the defect operators) 
\ie
\vev{\cO^{a;x}_m \, \cO^{b;y}_n \, \cO^{c;z}_k}^w & ~\equiv~
\trip{.5}{1}{$\cO^{a;x}_m$}{$\cO^{b;y}_n$}{$\cO^{c;z}_k$}{$w$} \, .
\fe
We view the small discs as boundaries resulting from blowing up point-like bulk defect operators.
Since all the boundaries are connected in the sense of not being separated by TDLs, we can skip step 1a and perform step 1b twice (with no TDL in between, i.e.\ $c=\cI$) to get a sphere one-point function (accompanied by some factors), which is then rearranged by step 2 into 
\ie
\frac{\D_{m,n,k}}{\td_m^2} ~ \tetjfill{m}{m}{b}{c}{a}{m}{w}{x}{y}{z}
\fe
and can be evaluated purely using the data of $\cC$, $\cM$ and the module trace.

The two- and four-point function derivations are completely analogous, and are therefore omitted.
In summary, we have arrived at the following formulae:
\ie
\label{SahandDynamical}
\vev{\cO^{a;x}_m \, \cO^{b;y}_n} &\equiv \twopt{.5}{1}{$\cO^{a;x}_m$}{$\cO^{b;y}_n$} = \frac{\D_{a,\bar b} \D_{m,n}}{\td_m} ~~ \Thjfill{m}{a}{m}{x}{y} \, ,
\\
\vev{\cO^{a;x}_m \, \cO^{b;y}_n \, \cO^{c;z}_k}^w &\equiv
\trip{.5}{1}{$\cO^{a;x}_m$}{$\cO^{b;y}_n$}{$\cO^{c;z}_k$}{$w$}
= \frac{\D_{m,n,k}}{\td_m^2} ~ \tetjfill{m}{m}{b}{c}{a}{m}{w}{x}{y}{z}
\\
\vev{\cO^{a;x}_m \, \cO^{b;y}_n \, \cO^{c;z}_k \, \cO^{d;w}_\ell}^{u,v}_e &\equiv
\fourpt{.5}{\cO^{a;x}_m}{\cO^{b;y}_n}{\cO^{c;z}_k}{\cO^{d;w}_\ell}{e}{u}{v}
= \frac{\D_{m,n,k,\ell}}{\td_m^3} ~~ \fourfill{1}{x}{y}{z}{w}{e}{u}{v}{m}{m}{m}{m}{a}{b}{c}{d}
\, .
\fe

\subsubsection*{Crossing symmetry}

Crossing symmetry is the statement that first, four-point functions can be built from three-point functions via cutting-and-sewing,
\ie
\label{FourPtCuttingSewing}
\vev{\cO^{a;x}_m \, \cO^{b;y}_n \, \cO^{c;z}_k \, \cO^{d;w}_\ell}^{u,v}_e = \sum_{\cU, \cV \in \cH^e} \vev{\cO^{a;x}_m \, \cO^{b;y}_n \, \cU^\dag}^u \times N^{-1}(\cU, \cV) \times \vev{\cV \, \cO^{c;z}_k \, \cO^{d;w}_\ell}^v \, ,
\fe
and second,
\ie
\vev{\cO^{a;x}_m \, \cO^{b;y}_n \, \cO^{c;z}_k \, \cO^{d;w}_\ell}^{u,v}_e
=
\sum_{f,s,t}
(F^{a,b,c}_{\bar d;e,f})^{u,v}_{s,t} \,
\vev{\cO^{b;y}_n \, \cO^{c;z}_k \, \cO^{d;w}_\ell \, \cO^{a;x}_m}^{s,t}_f \, .
\fe
The second condition is automatic according to our formula on the last line of \eqref{SahandDynamical}, together with the symmetric property \eqref{TracePairing} of the module trace.

We now present the proof of the first condition.
Let us regard the four-point function (up to factors of $\td_m$) as the module trace of a $\Hom_\cM(m,m)$, and perform an F-move between $e$ and $m$, 
\ie
\begin{gathered}
\begin{tikzpicture}
\fill [fill=gray!20] (0,0) rectangle (1,7);
\draw [line,->-=.52] (0,0) -- (0,3.5) node [right] {$m$} -- (0,7);
\draw [line,->-=.55] (0,2) ++(270:1) arc (270:180:1);
\draw [line,->-=.55] (0,2) -- (-.5,2) node [below] {$b$} -- (-1,2);
\draw [line,->-=.55] (0,5) ++(90:1) arc (90:180:1);
\draw [line,->-=.55] (0,5) -- (-.5,5) node [above] {$c$} -- (-1,5);
\draw [line,->-=.55] (-1,2) -- (-1,3.5) node [left] {$e$} -- (-1,5);
\node at (-1,1) {$a$};
\node at (-1,6) {$d$};
\draw (-1,2) node [left] {\scriptsize $u$};
\draw (-1,5) node [left] {\scriptsize $v$};
\draw (0,1) node [right] {\scriptsize $x$};
\draw (0,2) node [right] {\scriptsize $y$};
\draw (0,5) node [right] {\scriptsize $z$};
\draw (0,6) node [right] {\scriptsize $w$};
\end{tikzpicture}
\end{gathered}
\quad=\quad
\sum_{m', s, t}
(\tF^{\bar e, e, m}_{m; \cI, m'})^{u, v}
\quad
\begin{gathered}
\begin{tikzpicture}
\fill [fill=gray!20] (0,0) rectangle (1,7);
\draw [line,->-=.52] (0,0) -- (0,3.5) node [right] {$m'$} -- (0,7);
\draw [line,->-=.27,->-=.8] (0,2) ++(270:1) arc (270:90:1);
\draw [line,->-=.55] (0,2) -- (-.5,2) node [below] {$b$} -- (-1,2);
\draw [line,->-=.27,-<-=.8] (0,5) ++(90:1) arc (90:270:1);
\draw [line,->-=.55] (0,5) -- (-.5,5) node [above] {$c$} -- (-1,5);
\node at (-1,1) {$a$};
\node at (-1,6) {$d$};
\node at (-1,3) {$e$};
\node at (-1,4) {$e$};
\draw (0,3) node [right] {\scriptsize $u$};
\draw (0,4) node [right] {\scriptsize $v$};
\draw (0,1) node [right] {\scriptsize $x$};
\draw (0,2) node [right] {\scriptsize $y$};
\draw (0,5) node [right] {\scriptsize $z$};
\draw (0,6) node [right] {\scriptsize $w$};
\end{tikzpicture}
\end{gathered}
\, .
\fe
The simplicity of $m$ forces $m' = m$, and moreover, just like in our derivation of the boundary crossing relation in section~\ref{Sec:BoundaryCrossing}, the module trace factorizes,
\ie
\fourfill{1}{x}{y}{z}{w}{e}{u}{v}{m}{m}{m}{m}{a}{b}{c}{d}
~=
\sum_{u, v}
~
\frac{(\tF^{\bar e, e, m}_{m; \cI, m})^{u, v}}{\td_m}
~
\tetjfill{m}{m}{b}{\bar e}{a}{m}{w}{x}{y}{u}
~~
\tetjfill{m}{m}{d}{e}{c}{m}{w}{z}{w}{v} \, .
\fe
By relating $\cU^\dag = \cO^{\bar e;u}_{m,m}$ and $\cV = \cO^{e;v}_{m,m}$, and using the generalized boundary crossing relation \eqref{GeneralizedBoundaryCrossing}, we have proven \eqref{FourPtCuttingSewing}.

\subsubsection*{Lassos and correlators on general surfaces}

A general observable in a closed $\cC$-symmetric TFT is the expectation value of a network of TDLs and point-like bulk defect operators on a closed oriented Riemann surface.
Sewing theorems guarantee that any such observable can be built from the sphere three-point functions and the lassos, to be defined shortly, as long as the sphere four-point crossing and the torus one-point modular invariance are satisfied.

A lasso is a two-point function of defect operators separated by TDLs.
More precisely, consider two defect operators $\dO{m}{a;x}\in \cH^a$ and $\dO{n}{b;y} \in \cH^b$ that are separated by TDLs $c$ and $d$ as shown below (the TDL types attached to the defect operators are implicit in the superscripts)
\ie
\lasso{1}{$\dO{m}{a;x}$}{$d$}{$c$}{$\dO{n}{b;y}$}{$u$}{$v$} \,.
\fe

Now we can derive a formula for the lasso.
As before, we view the small discs as boundaries resulting from blowing up point-like bulk defect operators.
Then using the boundary crossing relation \eqref{GeneralizedBoundaryCrossing}, we can merge the two disc boundaries and rewrite the lasso in terms of a disk correlator 
\ie
\label{SahandLasso}
\lassoS{1}{$\dO{m}{a;x}$}{$d$}{$c$}{$\dO{n}{b;y}$}{$u$}{$v$} = \sum_{z,t} G_{zt} ~\times~ \fourfill{1}{x}{z}{y}{t}{c}{u}{v}{m}{n}{n}{m}{a}{d}{b}{\bar d} \, ,
\fe
where $z \in \cH^d_{m,n}$ and $t \in \cH^{\bar{d}}_{n,m}$ sums over appropriate bases of the Hilbert spaces, and $G$ is the inverse two-point function defined in \eqref{g.inverse}.

\subsection{Summary}
\label{sec:summary}

To recap, any observable in a $\cC$-symmetric TFT can be constructed from a set of defining data and performing cutting-and-sewing.
In the open sector, the disc correlators can almost be identified with the module category structure and the trace.
In the closed sector, the defining data consists of the one-point, two-point, and three-point functions of defect operators on the sphere, together with the lasso two-point functions.\footnote{The one- and two-point functions are tied to the normalization of $\vev{~}$ and the normalization of the point-like operators.
Hence, under a fixed normalization, the defining data consists of just the three-point functions and the lasso two-point functions.}
In the above, they have been completely determined in terms of module categorical data.
For convenience, we summarize the formulae below.

We remind the reader that $a, b, c, \dotsc$ label simple TDLs, $m, n, k, \dotsc$ label simple BCs, with $\td$ the quantum dimension, $x, y, z, \dotsc$ are basis junction vectors,\footnote{Recall that a gauge choice fixes a set of basis junction vectors.} 
and finally $\cO^{a;x}_m$ denotes the point-like defect operator in the Hilbert space $\cH^a$ created by a hollow disc boundary with BC $m$ and junction vector $x$.
\begin{flalign}
  &\textbf{1-point}:  && \vev{\cO_m}  = {\td_m}\,,&&\nonumber\\
  &\textbf{2-point}:  && \Big \langle \cO^{a;x}_m \, \cO^{\bar{a};y}_n \Big \rangle = \frac{\D_{m,n}}{\td_m} ~~ \Thjfill{m}{a}{m}{y}{x} \,,&&\nonumber\\
  &\textbf{3-point}:  && { \Big \langle \cO^{a;x}_m \, \cO^{b;y}_n \, \cO^{c;z}_k \Big \rangle }^w  = \frac{\D_{m,n,k}}{\td_m^2} ~ \tetjfill{m}{m}{b}{c}{a}{m}{w}{x}{y}{z} \,,&& \label{summary} \\
  &\textbf{Lasso}:  &&  \Bigg \langle \lasso{1}{$\dO{m}{a;x}$}{$d$}{$c$}{$\dO{n}{b;y}$}{$u$}{$v$} \Bigg \rangle  = \sum_{z,t} G_{zt} \left( \fourfill{1}{x}{z}{y}{t}{c}{u}{v}{m}{n}{n}{m}{a}{d}{b}{\bar d} \right) \,, &&\nonumber\\
    &&& \left( G^{-1} \right)^{yx} 
    = \Thjfill{n}{d}{m}{y}{x}~.&&\nonumber
\end{flalign}

If we interpret the expressions on the righthand side not as abstract traces of morphisms in a module category, but as physical disc correlators in the open sector, then the general philosophy of the open sector's completely determining the closed sector cannot be more obvious.
This not only echoes the sentiments of \cite{Moore:2006dw} in the context of open/closed TFT, but also appears to be a simplified version of the structure \cite{Fjelstad:2006aw} in rational conformal field theory.

\section{Regular topological field theories and examples}
\label{Sec:Examples}

For any fusion category $\cC$, there is a canonical $\cC$-symmetric TFT known as the regular TFT, where the category $\cM$ of BCs is isomorphic to $\cC$ itself.\footnote{Since $\cC$ acts on itself by fusion, it can be thought as a module category over itself and is known as the regular module category.
} 
There is a sense in which once we understand regular TFTs, we understand all TFTs.
Namely, as will be explained in section \ref{Sec:gauging}, instead of applying our general formulae \eqref{summary} to non-regular module categories, we could also start from a regular TFT for a different fusion category $\cC'$, and perform generalized gauging.

In the following, we first summarize our general recipe to construct regular TFTs.
Then we explicitly construct the regular TFTs for several fusion categories.
The {defining data} of some of these examples were previously obtained by bootstrap \cite{Chang:2018iay,Huang:2021ytb}.
However, the natural bootstrap convention is where the expectation value of the identity local operator $\one$ is normalized to one, the action of TDLs on local operators is as diagonal as possible, and the point-like defect operators have orthonormal two-point functions.
We use $\vvev{~}$ to denote the expectation value in the bootstrap normalization.
In the following, we adopt the notation of \cite{Chang:2018iay,Huang:2021ytb} as much as possible to make our results directly comparable.\footnote{However, here we denote the unit TDL by $\cI$ instead of $I$.}

\subsection*{Recipe}

The observables of a regular TFT can be constrcuted from the defining data whose formulae are summarized in section \ref{sec:summary}.
In the case of the regular TFT, the category of boundary condition is the same as $\cC$ itself.
Therefore, simple TDLs $a,b,c,\dots$ and simple BCs $m,n,k,\dots$ are both labeled by simple objects in $\cC$.
Moreover, the boundary quantum dimensions coincide with defect quantum dimensions; thus we set $\td_m = d_m$ in \eqref{summary}.

Note that for the case of the regular TFT, the righthand sides of the equations in \eqref{summary} become just diagrams in the fusion category $\cC$.
Therefore, one can evaluate them by the F-moves of $\cC$, and construct the TFT purely from the fusion categorical data.

\subsection{Fibonacci fusion category}

We denote the simple TDLs in the Fibonacci fusion category by $\cI$ and $W$, so that the nontrivial fusion rule and quantum dimension are
\ie
W^2 = \cI + W \, , \quad d_W = \zeta \equiv \frac{1+\sqrt5}{2} \, .
\fe
Moreover, we work in a gauge where the nontrivial F-symbols are
\ie
F^{W,W,W}_W =
\begin{pmatrix}
\zeta^{-1} & \zeta^{-1}
\\
1 & - \zeta^{-1}
\end{pmatrix} \, ,
\fe
such that
\ie
\Thun{.5} \ = \ \zeta^2 \, , \quad \tetun{.5} \ = \ - \zeta^2 \, , \quad \pentun{.33} \ = \ \zeta^2 \, .
\fe

In the regular TFT, the closed Hilbert spaces are spanned by the point-like operators
\ie
\cH~~ &: \quad \bO{{}\cI}, ~~ \bO{{}\cW} \, ,
\\
\cH^\cW &: \quad \dO{{}\cW}{\cW} \, ,
\fe
and the TDL $W$ acts on the local operators in $\cH$ by
\ie
\widehat\cW \cdot \cO_{\cI} = \cO_{\cI} \, , \quad \widehat\cW \cdot \cO_{\cW} = \cO_{\cI} + \cO_{\cW} \, .
\fe
The bulk Frobenius algebra \eqref{Frobenius} is
\ie
\bO{{}\cI} \bO{{}\cI} = \bO{{}\cI} \, , \quad \bO{{}\cW} \bO{{}\cW} = \frac{1}{\zeta} \bO{{}\cW} \, , \quad \bO{{}\cI} \bO{{}\cW} = 0 \, , \quad \vev{\bO{{}\cI}} = 1 \, , \quad \vev{\bO{{}\cW}} = \zeta \, ,
\fe
while the rest of the {defining data} \eqref{SahandDynamical}, \eqref{SahandLasso} are
\ie
& \vev{ \dO{{}\cW}{\cW} \dO{{}\cW}{\cW} } = \zeta \, ,
\qquad &&
\vev{ \dO{{}\cW}{\cW} \dO{{}\cW}{\cW} \bO{{}\cW} } = \frac{1}{d_W^2} \,
\Thun{.5} = 1 \, ,
\\
& \vev{ \dO{{}\cW}{\cW} \dO{{}\cW}{\cW} \bO{{}\cI} } = 0 \, ,
\qquad &&
\vev{ \dO{{}\cW}{\cW} \dO{{}\cW}{\cW} \dO{{}\cW}{\cW} } = \frac{1}{d_W^2} \,
\tetun{.5}
= - 1 \, ,
\fe
and
\ie
\vev{ \bell{1}{$\dO{{}\cW}{\cW}$}{}{$\bO{{}\cI}$} } &= \frac{1}{d_W} ~ \Thun{.5} = \zeta \, ,
~~
\vev{ \bell{1}{$\dO{{}\cW}{\cW}$}{}{$\bO{{}\cW}$} } &= \frac{\tetun{.5}}{\Thun{.5}} = -1 \, ,
\\
\vev{ \lassoX{1}{$\dO{{}\cW}{\cW}$}{}{}{$\dO{{}\cW}{\cW}$} } &= \frac{\pentun{.33}}{\Thun{.5}} = 1 \, .
\fe

Let us now convert to the bootstrap convention, where the local operators $\one, v_x$ diagonalize the TDL action,
\ie
\widehat\cW \cdot \one = \zeta \, , \quad \widehat\cW \cdot v_x = - \zeta^{-1} v_x \, ,
\fe
and the point-like defect operator in $\cH^\cW$ is $v_\mu$.
The basis transformation is
\ie
& \one = \bO{{}\cI} + \zeta \bO{{}\cW} \, ,
\quad
v_x = \zeta \bO{{}\cI} - \bO{{}\cW} \, ,
\quad
v_\mu = \sqrt{\zeta+\zeta^{-1}} \, \dO{{}\cW}{\cW} = \sqrt[4]{5} \, \dO{{}\cW}{\cW} \, .
\fe
A straightforward calculation gives
\ie
& \vvev{v_x v_x v_x} = 1
\, ,
\qquad
\vvev{v_\m v_\m v_x} = - \zeta^{-1}
\fe
and
\ie
& \vvev{\tri{.5}{0}{$v_\m$}{$v_\m$}{$v_\m$}} = - \sqrt{\frac{3 \sqrt{5}}{2}-\frac{5}{2}} \, ,
~~
\vvev{\bell{1}{$v_\m$}{}{$v_x$}} = \sqrt[4]{5} \, ,
\\
& \vvev{\lassoX{1}{$v_\m$}{}{}{$v_\m$}} = \zeta^{-1} \, ,
\fe
reproducing the bootstrap result of \cite{Chang:2018iay}.

\subsection{Ising fusion category}

The Ising fusion category has an invertible TDL $\eta$ and a non-invertible TDL ${N}$, with the fusion rules and quantum dimensions
\begin{equation}
	\eta^2 = \cI \, , \quad \eta  {N} = {N} \eta = {N} \, , \quad {N}^2=\cI+\eta\, , \quad \vev{\eta} = \cI \, , \quad \vev{{N}} = \sqrt2 \, .
\end{equation}
We work in unitary gauge in which the nontrivial F-symbols are
\ie
& F^{{N},{N},{N}}_{N} = {1 \over \sqrt{2}} \begin{pmatrix}  1 & 1 \\ 1  & -1 \end{pmatrix} \, ,
\quad
F^{\eta,{N},\eta}_{N} = -1 \, .
\fe
In the following, solid lines represent $N$, and dashed lines represent $\eta$.
From the F-symbols, one can derive that
\ie
\Thdashed{.5}{}{}{} = \sqrt2 \, , \quad \tetdashed{.5} = -\sqrt2 \, .
\fe

In the regular TFT, the nonempty closed Hilbert spaces are spanned by the point-like operators
\begin{align}
	\mathcal{H}~~ &: \quad \dO{\cI}{} \, , ~~ \dO{\eta}{} \, , ~~ \dO{{N}}{} \,,  \\
	\mathcal{H}^\eta~ &: \quad \dO{{N}}{\eta} \, ,
\end{align}
and the TDLs act on the local operators in $\cH$ by
\ie
\widehat\eta \cdot \cO_\cI &= \cO_\eta \, , \quad & \widehat\eta \cdot \cO_\eta &= \cO_\cI \, , \quad & \widehat\eta \cdot \cO_{N} &= \cO_{N} \, ,
\\
\widehat{N} \cdot \cO_\cI &= \cO_{N} \, , \quad & \widehat{N} \cdot \cO_\eta &= \cO_{N} \, , \quad & \widehat{N} \cdot \cO_{N} &= \cO_\cI + \cO_\eta \, .
\fe
The bulk Frobenius algebra \eqref{Frobenius} is
\ie
\dO{\cI}{} \dO{\cI}{} &= \dO{\cI}{}\,, \quad \dO{\eta}{} \dO{\eta}{} = \dO{\eta}{}\,, \quad \dO{{N}}{} \dO{{N}}{} = \frac{1}{\sqrt{2}} \dO{{N}}{}\,,
\fe
while the rest of the {defining data} \eqref{SahandDynamical}, \eqref{SahandLasso} are
\ie
&	\langle \dO{{N}}{} \dO{{N}}{} \dO{{N}}{} \rangle = \frac{1}{d_{N}^2 } ~ \lev{.5}{} = \frac{1}{\sqrt{2}} \, , \quad \langle \dO{{N}}{\eta} \dO{{N}}{\eta} \dO{{N}}{} \rangle = \frac{1}{\langle {N} \rangle^2 } ~ \Thdashed{.5}{}{}{} = \frac{1}{\sqrt{2}} \, ,
\fe
and
\ie
\vev{\bell{0}{$\dO{{N}}{\eta}$}{}{$\bO{1}$}} &= ~ \frac{\Thdashed{.5}{}{}{}}{\lev{.5}{}}
= 1 \, ,
~~
\vev{\bell{0}{$\dO{{N}}{\eta}$}{}{$\bO{\eta}$}} = ~ \frac{\tetdashed{.5}}{\Thdashed{.5}{}{}{}}
= -1 \, , \\
\vev{\lassoDashed{0}{$\dO{{N}}{\eta}$}{}{}{$\dO{{N}}{\eta}$}} &= \raisebox{.15in}{``} \Thun{.5} \raisebox{.15in}{"} = 0 \, .
\fe

Let us now convert to the bootstrap convention, where the local operators $\one, v_\varepsilon, v_\sigma$ diagonalize the TDL action,
\ie
\widehat\eta \cdot \one &= - \one \, , \quad &
\widehat\eta \cdot v_\varepsilon &= v_\varepsilon \, , \quad &
\widehat\eta \cdot v_\sigma &= - v_\sigma \, ,
\\
\widehat{N} \cdot \one &= \sqrt2 \, , \quad &
\widehat{N} \cdot v_\varepsilon &= -\sqrt2 v_\varepsilon \, , \quad &
\widehat{N} \cdot v_\sigma &= 0 \, ,
\fe
and the point-like defect operator in $\cH^\eta$ is $v_\mu$.
The basis transformation is
\ie
	\one &= \bO{1} + \bO{\eta} +\sqrt{2}\bO{{N}} \,, \quad&
	v_\varepsilon &= \bO{1} + \bO{\eta} -\sqrt{2}\bO{{N}} \,,
	\\
	v_\sigma &= \sqrt{2}\bO{1} - \sqrt{2} \bO{\eta} \,,\quad&
	v_\mu &= 2\dO{{N}}{\eta} \,.
\fe
A straightforward calculation gives
\ie
	v_\varepsilon^2 &= \one \, , \quad& v_\sigma^2 &= \one + v_\varepsilon \, , \quad& v_\varepsilon v_\sigma &= v_\sigma \,, \\
	v_\mu v_\mu &= \one - v_\epsilon \, , \quad& v_\varepsilon v_\mu &= - v_\mu \, , \quad& v_\sigma v_\mu &= 0  \, ,
\fe
and
\ie
& \vvev{\bell{0}{$v_\mu$}{}{$v_\varepsilon$}} = 0 \, ,
~~
\vvev{\bell{0}{$v_\mu$}{}{$v_\sigma$}} = 1 \, ,
\\
& \vvev{\lassoDashed{0}{$v_\mu$}{}{}{$v_\mu$}} = 0 \, .
\fe

In fact, the lasso action of ${N}$ (up to a factor of $\sqrt2$) becomes a \emph{unitary symmetry operator} $U(N)$ acting on $\mathcal{H} \oplus \mathcal{H}^\eta$,
\begin{align}
	U(N) \cdot \one = \one \,,\quad U(N) \cdot v_\varepsilon = - v_\varepsilon\,,\quad U(N) \cdot v_\sigma = v_\mu\,,\quad U(N) \cdot v_\mu =  v_\sigma\,.
\end{align}
This can be generalized to any TDL in an arbitrary fusion category: every TDL $a$ gives rise to a unitary symmetry operator
\begin{equation}
    U(a)~:~\mathcal{H}^{a \otimes \bar a} \to  \mathcal{H}^{\bar a \otimes a}~.
\end{equation}

\subsection{Haagerup $\cH_3$ fusion category}

We label the simple TDLs in the Haagerup $\cH_3$ fusion category by $\cI,\alpha,\alpha^2,\rho,\alpha\rho,\alpha^2\rho$.
The nontrivial fusion rules 
are
\ie
\A^3 = 1 \, , \quad \A \rho = \rho \, \A^2 \, , \quad \rho^2 = \cI + \cZ \, , \quad \cZ \equiv \sum_{i=0}^2 \A^i \rho \, ,
\fe
and the quantum dimensions are
\ie
\{d_\cI, d_\alpha, d_{\alpha^2}, d_{\rho}, d_{\alpha\rho}, d_{\alpha^2\rho}\}=\{1, 1, 1, \zeta, \zeta, \zeta\} \, , \quad \zeta=\frac{3+\sqrt{13}}{2} \, .
\fe
In the gauge of \cite{Huang:2021ytb} for the F-symbols,\footnote{Note that the gauge used in \cite{Huang:2021ytb} is slightly different from the unitary gauge of \cite{Huang:2020lox}.
}
we can derive the identities
\ie
& \Thun{.5} ~\equiv~ \Th{\rho}{\rho}{\rho}
\ = \
\zeta^2
\, ,
\quad
\tetun{.5} ~\equiv~ \tet{\rho}{\rho}{\rho}{\rho}{\rho}{\rho}
\ = \
\frac{-29-8\sqrt{13}}{3} \, ,
\\
& \pentun{.33} ~\equiv~ \fourun{1}{}{}{}{}{\rho}{}{}{\rho}{\rho}{\rho}{\rho}{\rho}{\rho}{\rho}{\rho}
\ = \
\frac{307+85\sqrt{13}}{18} \, ,
\fe
which will be used in the following.

In the regular TFT, the closed Hilbert spaces are spanned by the point-like operators
\ie
\cH~~~~ &: \quad \bO{\cI}, ~~ \bO{\A}, ~~ \bO{\A^2}, ~~ \bO{\rho}, ~~ \bO{\A\rho}, ~~ \bO{\A^2\rho} \, ,
\\
\cH^\rho~~~  &:\quad \dO{\rho}{\rho}, ~~ \dO{\A\rho}{\rho}, ~~ \dO{\A^2\rho}{\rho} \, ,
\\
\cH^{\A\rho}~~  &:\quad \dO{\rho}{\A\rho}, ~~ \dO{\A\rho}{\A\rho}, ~~ \dO{\A^2\rho}{\A\rho} \, ,
\\
\cH^{\A^2\rho}~  &:\quad \dO{\rho}{\A^2\rho}, ~~ \dO{\A\rho}{\A^2\rho}, ~~ \dO{\A^2\rho}{\A^2\rho} \, ,
\fe
and the TDL $\rho$ acts on the local operators in $\cH$ by
\ie
\widehat\rho \cdot \bO{\cI} &= \bO{\rho} \, , \quad 
& \widehat\rho \cdot \bO{\rho}&=\bO{\cI}+\bO{\rho}+\bO{\A\rho}+\bO{\A^2\rho} \, ,
\\
\widehat\rho \cdot \bO{\A} &= \bO{\A\rho} \, , \quad 
&\widehat\rho \cdot \bO{\A\rho} &= \bO{\A}+\bO{\rho}+\bO{\A\rho}+\bO{\A^2\rho} \, ,
\\
\widehat\rho \cdot \bO{\A^2} &= \bO{\A^2\rho} \, , \quad 
& \widehat\rho \cdot \bO{\A^2\rho} &= \bO{\A^2}+\bO{\rho}+\bO{\A\rho}+\bO{\A^2\rho} \, .
\fe

We now convert to the bootstrap convention, where the local operators $1, v, u_1, \bar{u}_1, u_2, \bar{u}_2$ are defined such that
\ie
\widehat\rho \cdot 1 &= \zeta \, , \quad &
\widehat\rho \cdot v &= -\zeta v \, , \quad &
\widehat\rho \cdot u_i &= \bar u_i \, , \quad &
\widehat\rho \cdot \bar u_i &= u_i \, ,
\\
\widehat{\alpha} \cdot \one &= \one \, , \quad &
\widehat{\alpha} \cdot v &= v \, , \quad &
\widehat{\alpha} \cdot u_i &= \omega \, u_i \, , \quad &
\widehat{\alpha} \cdot \bar u_i &= \omega^2 \, \bar u_i \, ,
\fe
with $i = 1,2$ and $\omega=e^{\frac{2\pi i}{3}}$.
This basis almost diagonalizes the TDL action, except that $\widehat{\rho}$ acts as charge conjugation on $u_i$.
The basis transformation is given by
\ie
\cO_{\cI}&=\frac{1}{\mathrm{dim}(\cC)}\left(1+\zeta v+\sqrt[4]{13}\sqrt{\zeta}(u_2+\bar u_2)\right)\,,\\
\cO_{\A}&=\frac{1}{\mathrm{dim}(\cC)}\left(1+\zeta v+\sqrt[4]{13}\sqrt{\zeta}(\omega u_2+\omega^2\bar u_2)\right)\,,\\
\cO_{\A^2}&=\frac{1}{\mathrm{dim}(\cC)}\left(1+\zeta v+\sqrt[4]{13}\sqrt{\zeta}(\omega^2 u_2+\omega \bar u_2)\right)\,,\\
\cO_{\rho}&=\frac{1}{\mathrm{dim}(\cC)}\left(1-\zeta^{-1}v+\sqrt[4]{13}\frac{1}{\sqrt{\zeta}}(u_1+\bar u_1)\right)\,,\\
\cO_{\A\rho}&=\frac{1}{\mathrm{dim}(\cC)}\left(1-\zeta^{-1}v+\sqrt[4]{13}\frac{1}{\sqrt{\zeta}}(\omega u_1+\omega^2\bar u_1)\right)\,,\\
\cO_{\A^2\rho}&=\frac{1}{\mathrm{dim}(\cC)}\left(1-\zeta^{-1}v+\sqrt[4]{13}\frac{1}{\sqrt{\zeta}}(\omega^2 u_1+\omega \bar u_1)\right)\,,
\fe
where the global dimension is
\ie
\mathrm{dim}(\cC) = \frac{3}{2}\left(13+3\sqrt{13}\right)\,.
\fe
As a result, the bulk Frobenius algebra in the bootstrap basis is
\ie
& v \times v = 1 + 3 v \, , \quad u_1 \times \bar u_1 = 1 -\zeta^{-1} v \, , \quad u_2 \times \bar u_2 = 1 + \zeta v \, ,
\\
& u_1 \times \bar u_2 = 0 \, , \quad u_1 \times u_1 = \sqrt{1 + \zeta^{-2}} \, \bar u_1 \, , \quad u_2 \times u_2 = \sqrt{1 + \zeta^2} \, \bar u_2 \, ,
\fe
matching \cite{Huang:2021ytb}.

There is a large amount of data involving defect operators, so here we only match a subset that involves $\dO{\rho}{\rho}$.
To match the defect three-point function, consider the two- and three-point functions
\ie\label{o2}
\vev{\dO{\rho}{\rho}\dO{\rho}{\rho}}=\frac{1}{d_\rho} \, \Thun{.5}=\zeta \, , 
\quad
\vev{\dO{\rho}{\rho}\dO{\rho}{\rho}\dO{\rho}{\rho}}=\frac{1}{d_\rho^2} \, \tetun{.5}=\frac{-7-\sqrt{13}}{6} \, .
\fe
In the defect Hilbert space $\cH^\rho$, it is fortunate that the basis chosen in \cite{Huang:2021ytb} (up to some redefinition freedom) is proportional to $\dO{\rho}{\rho}, \, \dO{\rho}{\alpha\rho} \, , \dO{\rho}{\alpha^2\rho}$.
Following the notation there, we denote by $o_{11}$ the defect operator proportional to $\dO{\rho}{\rho}$ in the natural bootstrap normalization.
We can deduce from \eqref{o2} that 
$\dO{\rho}{\rho}$ and $o_{11}$ are related by
\ie\label{vmu}
o_{11} = \pm \sqrt{\frac{\mathrm{dim}(\cC)}{\zeta}}\dO{\rho}{\rho} \, ,
\fe
where the sign is mere convention.
Using \eqref{SahandDynamical} and \eqref{SahandLasso}, we compute
\ie
\vvev{\tri{.5}{0}{$o_{11}$}{$o_{11}$}{$o_{11}$}} &= \mp \left(-\sqrt{\frac{\mathrm{dim}(\cC)}{\zeta}}\right)^3\frac{1}{\mathrm{dim}(\cC)}\vev{\dO{\rho}{\rho} \dO{\rho}{\rho}\dO{\rho}{\rho}} = \mp \sqrt{\frac{17\sqrt{13} - 52}{3}} \, ,
\\
\vvev{
\lassonoarrow{.5}{$o_{11}$}{}{}{$o_{11}$}{}{}
} &= \left(-\sqrt{\frac{\mathrm{dim}(\cC)}{\zeta}}\right)^2\frac{1}{\mathrm{dim}(\cC)}\frac{\pentun{.33}}{\Thun{.5}}=\frac{-1+5\sqrt{13}}{18}\, ,
\fe
reproducing the bootstrap result of \cite{Huang:2021ytb} under the minus sign convention for \eqref{vmu}.

\section{Non-regular topological field theories from generalized gauging}
\label{Sec:gauging}

An ordinary finite group global symmetry is captured by invertible TDLs (in a so-called pointed fusion category).
In the absence of 't Hooft anomalies, such a symmetry can be gauged.
For non-invertible TDLs, there is a generalized notion of gauging, sometimes refered to as a generalized orbifold, which was introduced in \cite{Frohlich:2009gb,Carqueville:2012dk,Brunner:2013xna} and expanded in \cite{Bhardwaj:2017xup}.

In this section, we show how to get all non-regular TFTs from regular TFTs via generalized gauging.
We first introduce the notion of algebra objects that are central to the procedure, construct the gauged theory, and finally explain how every non-regular TFT is related to a regular TFT by generalized gauging.
The exposition follows \cite{Bhardwaj:2017xup} closely.

\subsection{Algebra objects}

To begin, it is instructive to reformulate the gauging of finite group symmetries in the language of TDLs.
To gauge a finite group symmetry, we first couple the theory to a flat background gauge field and then make the gauge field dynamical, i.e.\ sum over distinct configurations.
In the TDL language, coupling the theory to a flat background gauge field configuration is equivalent to inserting a network of simple TDLs, and making the gauge field dynamical amounts to 
summing over inequivalent networks.

In more precise terms, to gauge a non-anomalous finite group symmetry $G$, we do the following.
We first consider the TDL $\cA=\bigoplus_{g\in G} g$ obtained by taking the direct sum of all TDLs in $G$.
Inserting a \emph{fine-enough} trivalent mesh of $\cA$ onto the spacetime achieves summing over all background gauge field configurations.
Here, a fine-enough mesh is a graph that is dual to a given triangulation of the spacetime manifold.
Note that we also need to fix a choice of the junction vector $\mu \in \Hom_\cC(\cA \otimes \cA, \cA)$ at the trivalent vertices of the graph.
A consistent choice, satisfying conditions described below such as \eqref{assoc} and \eqref{sep}, exists if and only if $G$ is non-anomalous.
There may be multiple choices corresponding to different discrete torsion \cite{Vafa:1986wx} elements of $H^2(G,U(1))$, or equivalently corresponding to coupling to different two-dimensional $G$ symmetry-protected topological phases.

The above procedure can be generalized to arbitrary fusion categories.
To perform a generalized gauging, we need a choice of a (non-simple) TDL $\cA \in \cC$, and a trivalent junction vector $\mu \in \Hom_\cC(\cA \otimes \cA, \cA)$
\begin{equation}
\begin{gathered}
\begin{tikzpicture}[scale = 1]
\draw [line,-<-=.6] (0,1) node[left] {$\cA$} -- (0,0);
\draw [line,-<-=.6] (0,0)  -- (-.87,-.5) node[left] {$\cA$};
\draw [line,-<-=.6] (0,0) -- (.87,-.5) node[right] {$\cA$};
\draw [fill=black] (0,0) circle (0.05) node[above right] {\scriptsize $\mu$};
\end{tikzpicture}
\end{gathered}\, .
\end{equation}
For the gauging procedure to be consistent, we need to require that the gauging is independent of the chosen triangulation.
Any two triangulations of a two-dimensional surface are related by a sequence of two basic moves.
In terms of the dual graph of the triangulation, the two moves correspond to the fusion and bubble moves of the mesh, and the requirement that any observable is invariant under these moves translates to the following conditions:
\begin{align}
    \begin{gathered}
\begin{tikzpicture}[scale=.7]
\fill[fill=none] (-2.5,-2) rectangle (1.5,2);
\draw [line,->-=.56] (-1,0) node [left] {\scriptsize $\mu$}
-- (-.5,0) node [above] {$\cA$} -- (0,0);
\draw [line,-<-=.56] (-1,0) -- (-1.5,.87) node [above left=-3pt] {$\cA$};
\draw [line,-<-=.56] (-1,0) -- (-1.5,-.87) node [below left=-3pt] {$\cA$};
\draw [line,-<-=.56] (0,0) -- (.5,-.87) node [below right=-3pt] {$\cA$};
\draw [line,->-=.56] (0,0) node [right] {\scriptsize $\mu$}
-- (.5,.87) node [above right=-3pt] {$\cA$};
\draw [fill=black] (0,0) circle (0.05);
\draw [fill=black] (-1,0) circle (0.05);
\end{tikzpicture}
\end{gathered}
&=
\begin{gathered}
\begin{tikzpicture}[scale=.7]
\fill[fill=none] (-2,-2.5) rectangle (2,1.5);
\draw [line,->-=.6] (0,-1) node [below] {\scriptsize $\mu$}
-- (0,-.5) node [right] {$\cA$} -- (0,0);
\draw [line,-<-=.56] (0,-1) -- (.87,-1.5) node [below right=-3pt] {$\cA$};
\draw [line,-<-=.56] (0,-1) -- (-.87,-1.5) node [below left=-3pt] {$\cA$};
\draw [line,-<-=.56] (0,0) -- (-.87,.5) node [above left=-3pt] {$\cA$};
\draw [line,->-=.56] (0,0) node [above] {\scriptsize $\mu$} -- (.87,.5) node [above right=-3pt] {$\cA$};
\draw [fill=black] (0,0) circle (0.05);
\draw [fill=black] (0,-1) circle (0.05);
\end{tikzpicture}
\end{gathered}
     &\leftrightarrow \quad &F^{\cA,\cA,\cA}_{\cA;\cA,\cA} \cdot \ket{\mu} \otimes \ket{\mu} = \ket{\mu} \otimes \ket{\mu}~.
     \label{assoc}\\
\begin{gathered}
\begin{tikzpicture}[scale=.7]
\fill[fill=none] (-2,-2) rectangle (2,2);
\draw [line,->-=.03,-<-=.52] (0,0) circle (0.5);
\draw [line,->-=.6] (0,-1.5) node[right] {$\cA$} -- (0,-0.5) ;
\draw [line,->-=.6] (0,0.5) -- (0,1.5) node[right] {$\cA$};
\draw (0.5,0) node[right] {$\cA$};
\draw (-0.5,0) node[left] {$\cA$};
\draw [fill=black] (0,0.5) circle (0.05) node[above left] {\scriptsize  $\mu$};
\draw [fill=black] (0,-0.5) circle (0.05) node[right] {\scriptsize  $\mu^\dagger$};
\end{tikzpicture}
\end{gathered}
&=
\begin{gathered}
\begin{tikzpicture}[scale=.7]
\fill[fill=none] (-2,-2) rectangle (2,2);
\draw [line,->-=.55] (0,-1.5) -- (0,1.5) node[right] {$\cA$};
\end{tikzpicture}
\end{gathered}
&\leftrightarrow \quad &\mu \circ \mu^\dagger = 1_\cA \in  \Hom_\cC(\cA , \cA)~. \label{sep}
\end{align}
In the {categorical language}, $\cA$ along with $\mu$ satisfying \eqref{assoc} and \eqref{sep} is called an associative and separable algebra object in $\cC$.

To make the gauging consistent, one has to further require that this algebra has a unit $u \in \Hom_\cC(\cI , \cA)$ satisfying
\begin{equation}
    \begin{gathered}
    \begin{tikzpicture}[scale=1]
    \draw [line,->-=.6] (0,-1) node [right] {\scriptsize $\mu$} -- (0,-.5) node [right] {$\cA$} -- (0,0);
    \draw [line,->-=.6] (0,-2) -- (0,-1.5) node [right] {$\cA$} -- (0,-1);
    \draw [line,-<-=.56] (0,-1) -- (-.87,-1.5);
    \draw (-.6,-1.25) node [above] {$\cA$};
    \draw [fill=black] (0,-1) circle (0.05);
    \draw [fill=black] (-.87,-1.5) circle (0.05) node [below left] {\scriptsize $u$};
    \end{tikzpicture}
    \end{gathered} ~ = \quad
     \begin{gathered}
    \begin{tikzpicture}[scale=1]
    \draw [line,->-=.6] (0,-2) -- (0,-1) node [right] {$\cA$} -- (0,0);
    \end{tikzpicture}
    \end{gathered}~.
\end{equation}
This guarantees that the theory after gauging has a vacuum.
Requiring a few more subtle properties makes $(\cA, \mu)$ into what is called a \emph{symmetric special Frobenius algebra} \cite{kirillov2002q,muger2003subfactors,Fuchs:2002cm}.\footnote{Non-invertible topological defect lines can also be gauged in higher dimensions.
See~\cite{Kaidi:2021gbs} for a discussion in three dimensions, and~\cite{Gaiotto:2019xmp} for a more general discussion in the context of category theory.}
From now on, we follow common practice and drop the adjective ``symmetric special Frobenius'' and refer to a gaugeable algebra simply as an \emph{algebra object}.
We also often just write $A$ instead of the pair $(A, \mu)$.

Generally there can be multiple algebra objects in a given fusion category $\cC$.
In particular, there is always the trivial algebra object $\cA = \cI$, which corresponds to gauging nothing.
When $\cC = \mathrm{Vec}_G^\omega$ is a finite group symmetry, where $\omega \in H^3(G, U(1))$ captures the 't Hooft anomaly, gauging an algebra object corresponds to gauging a non-anomalous subgroup of $G$.
However, for non-invertible symmetries, gauging an algebra object is not to be thought of as ``gauging a fusion subcategory''; the latter notion does not make sense unless the fusion subcategory admits a fiber functor.

\subsection{Gauged theory}
\label{subsec.gauged.theory}

Generalized gauging is performed by inserting a fine-enough mesh of an algebra object $\cA$ onto the spacetime.
More precisely, observables of the gauged theory can be computed from the observables of the ungauged theory in the presence of the mesh.
In particular, the torus partition function of the gauged theory $\cT' := \cT/A$ is given by the torus partition function of the ungauged theory $\cT$ as follows:
\begin{equation}
    Z_{\cT'}\left(~
        \begin{gathered}
        \begin{tikzpicture}[scale = 1]
        \draw [bg] (0,0) -- (2,0) -- (2,2) -- (0,2) -- (0,0);
        \end{tikzpicture}
        \end{gathered}
    ~\right)
    =
    Z_{\cT}\left(~
        \begin{gathered}
        \begin{tikzpicture}[scale = 1]
        \draw [bg] (0,0) -- (2,0) -- (2,2) -- (0,2) -- (0,0);
        \draw [line, ->-=0.56, ->-=0.15, ->-=0.95] (1,0) -- (1,1.05) node[right=-2pt] {$\cA$} -- (1,2);
        \draw [line, ->-=0.6] (0,1) -- (0.5,1.25) node[above] {$\cA$} -- (1,1.5);
        \draw [line, ->-=0.6] (1,0.5) -- (1.55,0.8) node[below] {$\cA$} -- (2,1);
        \draw [fill=black] (1,1.5) circle (0.05) node [right] {\scriptsize $\mu$};
        \draw [fill=black] (1,0.5) circle (0.05) node [left] {\scriptsize $\mu^\dagger$};
        \end{tikzpicture}
        \end{gathered}
    ~\right)~.
\end{equation}

To construct the gauged theory $\cT'$, we need a recipe for its defining data.
The intuitive guilding principle is that the mesh must be ``invisible'' to the new physical items in $\cT'$.
Roughly speaking, we demand the following conditions:
\begin{enumerate}
    \item $\cA$ can end on the new physical items. \label{cond.1}
    \item The correlation functions do not depend on the details of the mesh as long as the mesh is fine enough. \label{cond.2}
\end{enumerate}
Guided by these conditions, below we construct the local operators, BCs and TDLs in the gauged theory $\cT'$.

\subsubsection*{Local operators of the gauged theory}

A local operator in the gauged theory $\cT'$ corresponds to a point-like defect operator in $\cT$ living at the end of $\cA$,
\begin{equation}
    \begin{gathered}
    \begin{tikzpicture}[scale=.5,rotate=270]
    \draw [fill=black] (0.6,0) circle (.1);
    \draw [line, -<-=.5] (-2,0) -- (0.6,0);
    \draw (-1.5,0) node [left] {$\cA$};
    \draw (0.6,0) node [right] {$\cO^{\cA;x}_m$};
    \end{tikzpicture}
    \end{gathered}~, \qquad \cO^{\cA;x}_m \in \cH^\cA~.
\end{equation}
But not all such operators should be allowed due to condition~\ref{cond.2}; we require
\begin{equation}
\begin{gathered}
    \begin{tikzpicture}[scale=.5,rotate=270]
    \draw [fill=black] (0.6,0) circle (.1);
    \draw [line, -<-=.5, -<-=.17, -<-=.84] (-3,0) -- (0.6,0);
    \draw (-3,0) node [right] {$\cA$};
    \draw (0.6,0) node [right] {$\cO^{\cA;x}_m$};
    \draw [scale=1,domain=-3.141:3.141,smooth,variable=\t]
        plot ({(1.9+0.2*\t)*cos(\t r)+0.6},{(1.9+0.2*\t)*sin(\t r)});
    \draw [line,->-=1] (2.5,-.1) -- (2.5,.1) node [below] {$\cA$};
    \draw [fill=black] (-0.672,0) circle (.1);
    \draw (-0.672,0) node [right] {\scriptsize $\mu^\dagger$};
    \draw [fill=black] (-1.928,0) circle (.1);
    \draw (-1.928,0) node [left] {\scriptsize $\mu$};
    \end{tikzpicture}
\end{gathered} \quad = \quad
\begin{gathered}
    \begin{tikzpicture}[baseline=-36,scale=.5,rotate=270]
    \draw [fill=black] (0.6,0) circle (.1);
    \draw [line, -<-=.5] (-3,0) -- (0.6,0);
    \draw (-3,0) node [right] {$\cA$};
    \draw (0.6,0) node [right] {$\cO^{\cA;x}_m$};
    \end{tikzpicture}
\end{gathered} \, , \label{invariant.states}
\end{equation}
which means that only the operators invariant under $\cA$ are kept.
Note that in the usual orbifolding of a finite group $G$, one adds $G$-twisted sectors and projects to the $G$-invariant states.
Adding twisted sectors corresponds to considering $\cH^\cA$, and projecting to the invariant states corresponds to keeping the operators that satisfy \eqref{invariant.states}.

\subsubsection*{BCs of the gauged theory}

A BC of the gauged theory is given by a (not-necessarily-simple) BC $M \in \cM$ such that the mesh can end on it freely and consistently.\footnote{Since lowercase is reserved for simple objects, here we use the uppercase $M$ to stress that it can be non-simple.
Similarly, the not-necessarily-simple TDL below is denoted by $C$.
}
Condition~\ref{cond.1} requires the existence of a junction $x\in \mathrm{Hom}_\cM(\cA \otimes M, M)$ such that $\cA$ can end on $M$,
\begin{equation}
\begin{gathered}
\begin{tikzpicture}[scale=1]
\fill [fill=gray!20] (0,-1) rectangle (1,1);
\draw [line,->-=.55] (0,-1) -- (0,0);
\draw (0,-0.6) node [right] {$M$};
\draw [line,->-=.55] (0,0) -- (0,1);
\draw (0,0.6) node [right] {$M$};
\draw [line,-<-=.55] (0,0) -- (-1,0);
\draw (-0.6,0) node [below] {$\cA$};
\filldraw[black] (0,0) circle (1pt) node[right] {\scriptsize $x$};
\end{tikzpicture}
\end{gathered} \, ,
\end{equation}
and condition~\ref{cond.2} requires the following consistency:
\begin{equation}
\begin{gathered}
\begin{tikzpicture}[scale=1]
\fill[fill=gray!20] (0,0)--(.5,-.87)--(.5,.87);
\draw [line,->-=.6] (-1,0) node [left] {\scriptsize $\mu$}
-- (-.5,0) node [above] {$\cA$} -- (0,0);
\draw [line,-<-=.56] (-1,0) -- (-1.5,.87) node [above left=-3pt] {$\cA$};
\draw [line,-<-=.56] (-1,0) -- (-1.5,-.87) node [below left=-3pt] {$\cA$};
\draw [line,-<-=.56] (0,0) -- (.5,-.87) node [below right=-3pt] {$M$};
\draw [line,->-=.56] (0,0) node [right] {\scriptsize $x$}
-- (.5,.87) node [above right=-3pt] {$M$};
\end{tikzpicture}
\end{gathered}
~~=~~
\begin{gathered}
\begin{tikzpicture}[scale=1]
\fill[fill=gray!20] (.87,-1.5)--(0,-1)--(0,0)--(.87,.5);
\draw [line,->-=.56] (0,-1) node [below] {\scriptsize $x$}
-- (0,-.5) node [right] {$M$} -- (0,0);
\draw [line,-<-=.56] (0,-1) -- (.87,-1.5) node [below right=-3pt] {$M$};
\draw [line,-<-=.56] (0,-1) -- (-.87,-1.5) node [below left=-3pt] {$\cA$};
\draw [line,-<-=.56] (0,0) -- (-.87,.5) node [above left=-3pt] {$\cA$};
\draw [line,->-=.56] (0,0) node [above] {\scriptsize $x$}
-- (.87,.5) node [above right=-3pt] {$M$};
\end{tikzpicture}
\end{gathered}\, , \qquad
\begin{gathered}
\begin{tikzpicture}[scale=1]
\fill [fill=gray!20] (0,-1) rectangle (1,1);
\draw [line,->-=.55] (0,-1) -- (0,0);
\draw (0,-0.6) node [right] {$M$};
\draw [line,->-=.55] (0,0) -- (0,1);
\draw (0,0.6) node [right] {$M$};
\draw [line,-<-=.55] (0,0) -- (-1,0);
\draw (-0.6,0) node [below] {$\cA$};
\filldraw[black] (-1,0) circle (1pt) node[below] {\scriptsize $u$};
\filldraw[black] (0,0) circle (1pt) node[right] {\scriptsize $x$};
\end{tikzpicture}
\end{gathered}
\quad=\quad
\begin{gathered}
\begin{tikzpicture}[scale=1]
\fill [fill=gray!20] (0,-1) rectangle (1,1);
\draw [line,->-=.55] (0,-1) -- (0,1);
\draw (0,0) node [right] {$M$};
\end{tikzpicture}
\end{gathered} \, .
\label{A.module.associativity}
\end{equation}
A pair $(M,x)$ satisfying \eqref{A.module.associativity} is called a left $\cA$-module in $\cM$.
We have learned that the BCs of the gauged theory are given by the category of left $\cA$-modules in $\cM$, denoted ${}_\cA \cM$.

\subsubsection*{TDLs of the gauged theory}

The construction of the TDLs of the gauged theory is similar to that of the BCs, except that the mesh can now end both from the left and from the right.
Therefore, the TDLs of the gauged theory correspond to the category of $\cA$-bimodules in $\cC$.
An $\cA$-bimodule is a (not-necessarily-simple) TDL $C \in \cC$
with left and right junctions $x_L \in \mathrm{Hom}_\cC(\cA \otimes C, C)$ and $x_R \in \mathrm{Hom}_\cC(C \otimes \cA, C)$ satisfying
\begin{equation}
\begin{gathered}
\begin{tikzpicture}[scale=1,rotate=180]
\draw [line,-<-=.55] (0,-1) -- (0,0);
\draw (0,-0.4) node [right=-2pt] {$C$};
\draw [line,-<-=.55] (0,0) -- (0,1);
\draw (0,0.8) node [right=-2pt] {$C$};
\draw [line,-<-=.55] (0,0) -- (-1,1);
\draw (-0.6,0.4) node [above] {$\cA$};
\filldraw[black] (0,0) circle (1pt) node[left] {\scriptsize $x_R$};
\draw [line,->-=.55] (0,-1) -- (0,-2);
\draw (0,-1.8) node [right=-2pt] {$C$};
\draw [line,-<-=.55] (0,-1) -- (2,1);
\draw (1.1,0) node [above] {$\cA$};
\filldraw[black] (0,-1) circle (1pt) node[right] {\scriptsize  $x_L$};
\end{tikzpicture}
\end{gathered}
\quad=\quad
\begin{gathered}
\begin{tikzpicture}[scale=1]
\draw [line,->-=.55] (0,-1) -- (0,0);
\draw (0,-0.8) node [left=-2pt] {$C$};
\draw [line,->-=.55] (0,0) -- (0,1);
\draw (0,0.5) node [left=-2pt] {$C$};
\draw [line,-<-=.55] (0,0) -- (-1,-1);
\draw (-0.6,-0.4) node [above] {$\cA$};
\filldraw[black] (0,0) circle (1pt) node[right] {\scriptsize  $x_L$};
\draw [line,->-=.55] (0,1) -- (0,2);
\draw (0,1.8) node [left=-2pt] {$C$};
\draw [line,-<-=.55] (0,1) -- (2,-1);
\draw (1.1,0) node [above] {$\cA$};
\filldraw[black] (0,1) circle (1pt) node[left] {\scriptsize  $x_R$};
\end{tikzpicture}
\end{gathered}
\end{equation}
and the left and right associativity conditions analogous to \eqref{A.module.associativity}.
The category of $\cA$-bimodules in $\cC$ is commonly denoted by ${}_{\cA}\cC_\cA$,\footnote{Note that ${}_{\cA}\cC_\cA$ is a fusion category when $\cC$ is a fusion category and $\cA$ an \emph{indecomposable} algebra.
Generally, ${}_{\cA}\cC_\cA$ is a multifusion category.} and is called the dual or quantum symmetry.\footnote{Just like $\cC$ was not the set of all TDLs to begin with in the ungauged theory, here we are not presenting the recipe for \emph{all} TDLs in the gauged theory.
We are simply asking what happens to the TDLs in $\cC$ after gauging.
}

Thus we have learned that the gauged theory generally possess a dual ${}_{\cA}\cC_\cA$ symmetry, which is a generalization of the well-known fact that the $\bZ_N$ orbifold of a two-dimensional QFT has a dual $\widetilde{\bZ}_N$ symmetry \cite{Vafa:1989ih}.
In the $\bZ_N$ case, it is also well-known that gauging the dual $\widetilde{\bZ}_N$ symmetry recovers the original theory; in other words, gauging finite abelian symmetries is an invertible operation on the space of two-dimensional QFTs.
The situation for fusion category symmetries is completely parallel:
there is a dual algebra object $\cA' = \bar{\cA} \otimes \cA$ in $\cC'={}_{\cA}\cC_\cA$ such that ${}_{\cA'}{\cC'}_{\cA'} = \cC$, and the gauging of $\cA'$ in $\cC'$ is the inverse operation of gauging $\cA$ in $\cC$.

\subsection{Non-regular topological field theories and dual symmetry}

To explain why all non-regular TFTs 
can be obtained from regular TFTs, we invoke the following theorem:

{\it Given a fusion category $\cC$, there is one-to-one correspondence between distinct ways of gauging $\cC$ and module categories over $\cC$~\cite{ostrik2003module,Bhardwaj:2017xup}.
More precisely, for any algebra object $\cA$ in $\cC$, the category of right $\cA$-modules in $\cC$ denoted by $\cC_\cA$ is a left module category over $\cC$.
Moreover, any $\cC$-module category $\CM$ can be realized as $\cC_\cA$ for some algebra object $\cA$.}\footnote{Such an algebra object may not be unique.
Two algebra objects leading to the same module category are called Morita equivalent.
Moreover, $\cA$ is indecomposable if and only if $\cC_\cA$ is an indecomposable $\cC$-module category.
}

Suppose we want to construct a $\cC$-symmetric TFT whose BCs are described by a module category $\CM$.
The above theorem tells us that there exists an algebra object $\cA$ such that $\CM = \cC_\cA$.
Now, take the dual fusion category $\cC'={}_{\cA}\cC_\cA$, and consider the regular $\cC'$-symmetric TFT whose category of BCs is isomorphic to $\cC'$.
If we gauge the dual algebra object $\cA' = \bar{\cA} \otimes \cA$ in $\cC'$, then
the category of BCs 
becomes ${}_{\cA'} {\cC'}$.
But in fact ${}_{\cA'} {\cC'}$---which is a $\cC$-module category since $\cC = {}_{\cA'}{\cC'}_{\cA'}$---is the same as $\cC_\cA$.

The TFT constructed from a fusion category $\cC$ and a left module category $\CM = \cC_\cA$ always has an extra ${\cC'}^\text{op}$ symmetry, where the superscript op denotes orientation-reversing all TDLs, or equivalently taking the complex conjugation of all F-symbols.
The point is that a left module category $\cC_\cA$ over $\cC$ is at the same time a right module category ${}_{\cA'} \cC'$ over $\cC'$, and the two structures together form a $(\cC, \cC')$-bimodule category.
A natural physical picture for a bimodule category is that it captures an interface between a theory with $\cC$ symmetry on the left and one with $\cC'$ symmetry on the right.
However, we can fold along the interface to obtain a boundary with $\cC \boxtimes {\cC'}^\text{op}$ symmetry on the left and nothing on the right.
Restricting to the unit object $\cI' \in {\cC'}^\text{op}$ recovers the original TFT structure with only the $\cC$ symmetry, and restricting to the unit object $\cI \in \cC$ gives (the parity-reversal of) the gauged TFT with only the ${\cC'}^\text{op}$ symmetry.
Note that $\cC \boxtimes {\cC'}^\text{op}$ is generally not a faithful symmetry of the TFT.

\section*{Acknowledgements}

We are grateful to Kantaro Ohmori and Yuji Tachikawa for facilitating this collaboration, and to Yifan Wang for helpful discussions.
We also thank Zohar Komargodski, Kantaro Ohmori, Shu-Heng Shao, Yuji Tachikawa, and Yifan Wang for comments on the first draft.
YL thanks Noah Snyder and SS thanks Jin-Cheng Guu for consultation on the Frobenius-Schur indicator.
This material is based upon work supported by the U.S.
Department of Energy, Office of Science, Office of High Energy Physics, under Award Number DE-SC0011632.
YL is supported by the Simons Collaboration Grant on the Non-Perturbative Bootstrap.
SS is also supported in part by the Simons Foundation grant 488657 (Simons Collaboration on the Non-Perturbative Bootstrap).

\appendix

\section{Semisimplicity from unitarity/reflection-positivity}
\label{app.semisimple}

In this appendix, we prove that the category of \emph{all} TDLs and the category of BCs in a unitary/reflection-positive TFT are semisimple.

\subsection{Category of topological defect lines}

Let $\tC$ denote the category of \emph{all} TDLs and topological point-like defect operators in a unitary/reflection-positive TFT.
As was repeatedly emphasized in the main text, the category $\tC$ is generally different from the fusion category symmetry $\cC$.
We will show that $\tC$ is semisimple, i.e.\ any indecomposable TDL $a \in \tC$ is simple, meaning that $\Hom_\tC(a,a) \cong \bC$.
If $\tC$ is generated by finitely many TDLs, then what we will have proven is that $\tC$ is a multifusion category.
If it turns out that the unit TDL is simple, then $\tC$ is in fact a fusion category.

We proceed by first showing that $\Hom_\tC(a,a)$ forms a (not-necessarily-commutative) Frobenius algebra for any TDL $a$.
Then we use unitarity/reflection-positivity to prove that this Frobenius algebra is semisimple.
Then we are done, because any semisimple algebra has a complete set of idempotents (orthogonal projectors), and we can use them to project $a$ onto its simple components, thereby decomposing any TDL into simple TDLs.

\subsubsection*{The Frobenius algebra $\Hom_\tC(a,a)$}

To show that $\Hom_\tC(a,a)$ form a Frobenius algebra, we invoke the state-operator correspondence to identify $\Hom_\tC(a,a)$ with the defect Hilbert space on the circle twisted by $a \otimes \bar{a}$
\begin{equation}
    \Hom_\tC(a,a) \equiv \cH^{a \otimes \bar{a}}~.
\end{equation}
Defining a Frobenius algebra structure on $\cH^{a \otimes \bar{a}}$ amounts to defining the multiplication, unit, and trace.
Such structures on $\cH^{a \otimes \bar{a}}$ can be defined almost the same way as on $\cH$,
except that now we have to dress different two-dimensional cobordisms connecting collections of circles with the TDL $a \otimes \bar{a}$:
\begin{align}
&    \mu : \cH^{a \otimes \bar{a}} \times \cH^{a \otimes \bar{a}} \to \cH^{a \otimes \bar{a}} \quad 
\leftrightarrow \quad 
    \begin{gathered}
    \begin{tikzpicture}[scale=.25]
        \draw [bg] (0,0) ellipse (2 and 1);
        \draw [bg] (-3, -7)++(180:2 and 1) arc (180:360:2 and 1);
        \draw [bg, dashed] (-3, -7)++(0:2 and 1) arc (0:180:2 and 1);
        \draw [bg] (3, -7)++(180:2 and 1) arc (180:360:2 and 1);
        \draw [bg, dashed] (3, -7)++(0:2 and 1) arc (0:180:2 and 1);
        \draw [bg]   (0,-7)++(0:1) arc (0:180:1);
        \draw [bg] (-5,-7) .. controls (-5,-5) and (-2,-3) .. (-2,0);
        \draw [bg] (5,-7) .. controls (5,-5) and (2,-3) .. (2,0);
        \draw [line, -<-=.5] (-1,-0.866) -- (-4,-7.866);
        \draw [line, ->-=.5] (1,-0.866) -- (4,-7.866);
        \draw [line, -<-=.8] (-3,-8) -- (-1.2,-4.4);
        \draw [line, ->-=.8] (3,-8) -- (1.2,-4.4);
        \draw [line] (0,-5)++ (26.56:1.34) arc (26.56:153.43:1.34);
    \end{tikzpicture}
    \end{gathered}
    ~,\\
&     1 \in \cH^{a \otimes \bar{a}} \quad 
    \leftrightarrow \quad 
    \begin{gathered}
    \begin{tikzpicture}[scale=.25]
        \draw [bg] (0, 0)++(0:3 and 1) arc (0:360:3 and 1);
        \draw [bg] (0, 0)++(180:3) arc (180:360:3);
        \draw [line, ->-=.9] (1,-0.943) -- (1,-1.743);
        \draw [line, -<-=.9] (-1,-0.943) -- (-1,-1.743);
        \draw [line] (1,-1.743) to[out=270,in=270] (-1,-1.743);
    \end{tikzpicture}
    \end{gathered} \, ,
\qquad
    \theta: \cH^{a \otimes \bar{a}} \to \bC \quad 
    \leftrightarrow \quad
    \begin{gathered}
    \begin{tikzpicture}[scale=.25]
        \draw [bg, dashed] (0, 0)++(0:3 and 1) arc (0:180:3 and 1);
        \draw [bg] (0, 0)++(180:3 and 1) arc (180:360:3 and 1);
        \draw [bg] (0, 0)++(0:3) arc (0:180:3);
        \draw [line, -<-=.9] (1,-0.943) -- (1,-0.143);
        \draw [line, ->-=.9] (-1,-0.943) -- (-1,-0.143);
        \draw [line] (1,-0.143) to[out=90,in=90] (-1,-0.143);
    \end{tikzpicture}
    \end{gathered}
    ~.
\end{align}

Let us show that the bulk Frobenius algebra $\cH$, namely the $a=\cI$ case, is semisimple.
This was already proven in \cite{Durhuus:1993cq},\footnote{See \cite{Sawin:1995rh} for a discussion of similar statements in higher-dimensional TFTs.} but we repeat it here in order to to generalize it to the $a \neq \cI$ case.
It is well-known that a finite-dimensional algebra is semisimple if and only if the following trace in the regular representation is non-degenerate:\footnote{See \cite[footnote 11]{Moore:2006dw} or \cite[proposition 5.5.3]{cohn2012basic}.}
\begin{equation}
    \mathrm{tr}(\cO_1 \cO_2) := \sum_x \langle x | \cO_1 \cO_2 | x \rangle~.
\label{torus.2.pt}
\end{equation}
However, this trace is nothing but the torus two-point function, which is positive-definite by unitarity/reflection-positivity.
Hence the Frobenius algebra $\cH$ is semisimple.

For an arbitrary TDL $a$, we can again take the trace of the Frobenius algebra $\cH^{a \otimes \bar{a}}$ in its regular representation.
Similar to the $a=\cI$ case, this trace is a torus two-point function in the presence of $a \otimes \bar{a}$ wrapping a one-cycle
\begin{equation}
    \mathrm{tr}(\cO_1 \cO_2) = ~
    \begin{gathered}
    \begin{tikzpicture}[scale=1.1,baseline=30]
	\draw [line,->-=.3,->-=.8] (0,0) ellipse (1.15 and .55);
	\draw [line,-<-=.3,-<-=.8] (0,0) ellipse (1.35 and .75);
	\node[above] at (0,.7) {$\,a$};
	\node[above] at (0,-0.6) {$\,a$};
 	\draw [fill=black] (-1.35,0) circle (0.04) node[left=-2pt] {\scriptsize $\cO_1$};
 	\draw [fill=black] (1.35,0) circle (0.04) node[right=-2pt] {\scriptsize $\cO_2$};
	\draw (0,0) ellipse (1.8 and 1.1);
	\node[right] at (1.2,-1) {\scriptsize $\,T^2$};
	\begin{scope}[scale=.8]
		\path[rounded corners=24pt] (-.9,0)--(0,.6)--(.9,0) (-.9,0)--(0,-.56)--(.9,0);
		\draw[rounded corners=28pt] (-1.1,.1)--(0,-.6)--(1.1,.1);
		\draw[rounded corners=24pt] (-.9,0)--(0,.6)--(.9,0);
	\end{scope}
    \end{tikzpicture}
    \end{gathered} ~ .
\end{equation}
Such a torus two-point function is again positive-definite by unitarity/reflection-positivity, which finishes the proof that the Frobenius algbera $\cH^{a \otimes \bar{a}}$ is semisimple.

This proof can be easily extended to QFTs.
The only difference is that $\Hom_\tC(a,a)$ is not with the full defect Hilbert space $\cH^{a \otimes \bar{a}}$, but the topological subspace on which the Frobenius algebra structure can be defined.

An alternative proof for TFTs is as follows.
First note that any two-dimensional TFT has a state sum construction \cite{Fukuma:1993hy, Lauda:2006mn}, which can be interpreted as saying that a TFT with bulk Frobenius algebra $\cH$ can be obtained from a generalized gauging of the trivial theory.
The TDLs of the trivial theory are given by the category of finite-dimensional vector spaces, commonly denoted by $\mathrm{Vec}$.
Then $\cH$ can be thought of as an algebra object in $\mathrm{Vec}$, and we can gauge $\cH$ in the trivial theory to obtain the desired TFT with bulk Frobenius algebra $\cH$.
Based on the discussion in section \ref{subsec.gauged.theory}, $\tC$ is identified with the category of bimodules over $\cH$, commonly denoted by $_{\cH} \mathrm{Vec}_{\cH}$.
It is a theorem that if $\cH$ is semisimple, then its category of bimodules is also semisimple \cite{ostrik2003module}.

\subsection{Category of boundary conditions}

In the above, if we set $a=\cI$, then we find that the bulk Frobenius algebra is semisimple in unitary/reflection-positive theories.
But the category of BCs is nothing but the category of modules over the bulk Frobenius algebra \cite[theorem~2]{Moore:2006dw}.
Since the category of modules over a semisimple algebra is always semisimple \cite{ostrik2003module}, we have proven semisimplicity for BCs.

\section{Remarks on the orientation-reversal anomaly}
\label{App:FS}

We have assumed the triviality of Frobenius-Schur indicators throughout the main text to simplify the discussion.
This appendix provides some short remedial remarks.

Consider a self-dual TDL $a\simeq \bar{a}$, and choose an isomorphism
\ie
\begin{gathered}
\begin{tikzpicture}[scale=1]
\draw [line,-<-=.77,->-=.27] (0,-1) -- (0,-.5) node [right] {$a$} -- (0,0) \dtb{right}{0}{\scriptsize $\zeta$} -- (0,.5) node [right] {$a$} -- (0,1);
\end{tikzpicture}
\end{gathered} \, , \qquad \zeta \in \mathrm{Hom}(a,\bar{a}) \, .
\fe
The choice of $\zeta$ is not canonical, and relatedly we cannot always identify $a$ with $\bar{a}$.
Consider the dual junction $\zeta^\vee \in \mathrm{Hom}(a,\bar{a})$ given by a 180-degree rotation.\footnote{The dual on morphisms given by a rotation is a linear map, whereas the adjoint given by a reflection is an antilinear map.
One can think of the dual as the transpose, in the following sense.
For the representation category of a finite group, $\zeta$ is a similarity transformation between a representation $a$ and its complex conjugate representation $\bar a$.
The dual $\zeta^\vee$ corresponds to taking the transpose of the similarity matrix, and the adjoint $\zeta^\dag$ corresponds to taking the adjoint.
}
Since $\mathrm{Hom}(a,\bar{a})$ is one-dimensional for self-dual $a$, $\zeta^\vee$ is proportional to $\zeta$,
\begin{equation}
    \label{ZetaVee}
    \zeta^\vee = \chi_a \zeta\,,
\end{equation}
and the proportionality constant $\chi_a$ is known as the Frobenius-Schur indicator of $a$.
Note that $\chi_a$ does not depend on the choice of $\zeta$, and is an invariant of the isomorphism class of $a$.
Moreover, since $(\zeta^\vee)^\vee=\zeta$, the possible values for the Frobenius-Schur indicator are
\ie
\chi_a = \pm 1 \, .
\fe
In the case of $\chi_a = 1$, the junction $\zeta$ can be consistently trivialized, and one can ignore the arrows and truly identify $a$ with $\bar{a}$.
However, when $\chi_a = -1$, the junction is non-trivial, and we should keep track of the orientation on $a$.

The physical meaning of the Frobenius-Schur indicator in the context of two-dimensional QFT is as follows \cite{Chang:2018iay,Chang:2020aww}.
Any TDL comes with the freedom of an extrinsic curvature improvement term (ECIT).
Only for a specific choice of the ECIT coefficient is a TDL $a$ truly isotopic on a curved Riemann surface; otherwise, upon an isotopy deformation from path $\cP$ to $\cP' = \cP + \partial\cR$, the correlator changes by a phase proportional to the integral of the curvature over $\cR$, signifying an isotopy anomaly.\footnote{In the tensor categorical framework, the evaluation, coevaluation, and pivotal structures keep track of the ECIT as follows.
On a flat surface, when all TDLs are straight lines aligned along a single (say vertical) direction, the ECIT localizes at the points of folding.
Opposite phases arise from the ECIT under a left folding versus a right folding, and these phases are book-kept by the left and right (co)evaluation maps.
}
If the ECIT coefficients for $a$ and $\bar a$ can be chosen independently, then the isotopy anomaly can always be eliminated.
However, such a choice may forbid us from truly identifying $a$ with $\bar a$ even when $a$ is self-dual, signifying an orientation-reversal anomaly.
This happens precisely when the Frobenius-Schur indicator for $a$ is nontrivial.

Normally, one is free to choose whether to eliminate the orientation-reversal anomaly and suffer the isotopy anomaly, or the other way around.
However, since a topological field theory should not require a metric structure, it is more natural to forbid the the isotopy anomaly and live with the possible presence of an orientation-reversal anomaly.
Thus, if a self-dual TDL $a$ has a nontrivial Frobenius-Schur indicator, then $a, \bar a$ should be distinguished.
This is not to say that $a$ and $\bar a$ become independent simple objects, but just that we use the extra labeling to keep track of the orientation-reversal anomaly.
In particular, $\cH^a$ and $\cH^{\bar a}$ are still the same Hilbert space, just in different bases.

\bibliography{refs}
\bibliographystyle{JHEP}

\end{document}